\documentclass[a4paper,12pt,preprint,amsmath,showpacs,
nofootinbib,superscriptaddress]{revtex4}
\usepackage{mathrsfs}
\usepackage{}
\usepackage{amssymb}
\usepackage{amsmath}
\usepackage{graphicx}
\usepackage{multirow}
\usepackage{subfigure}
\usepackage{float}

\begin{document}

\title{Constraints on flavor-changing neutral-current $Htq$ couplings from the signal of $tH$ associated production  with  QCD next-to-leading order accuracy at the LHC}
\author{Yan  Wang}
\affiliation{Department of Physics and State Key Laboratory of
Nuclear Physics and Technology, Peking University, Beijing 100871,
China}
\author{Fa Peng Huang}
\affiliation{Department of Physics and State Key Laboratory of
Nuclear Physics and Technology, Peking University, Beijing 100871,
China}
\author{Chong Sheng Li}
\email{csli@pku.edu.cn}
\affiliation{Department of Physics and State Key Laboratory of
Nuclear Physics and Technology, Peking University, Beijing 100871,
China}
\affiliation{Center for High Energy Physics, Peking University, Beijing, 100871, China}

\author{Bo Hua Li}
\affiliation{Department of Physics and State Key Laboratory of
Nuclear Physics and Technology, Peking University, Beijing 100871,
China}
\author{Ding Yu Shao}
\affiliation{Department of Physics and State Key Laboratory of
Nuclear Physics and Technology, Peking University, Beijing 100871,
China}
\author{Jian Wang}
\affiliation{Department of Physics and State Key Laboratory of
Nuclear Physics and Technology, Peking University, Beijing 100871,
China}

\date{\today}


\begin{abstract}
We study a generic Higgs boson and a top quark  associated production via model-independent flavor-changing neutral-current  couplings at the LHC,
including  complete QCD next-to-leading order (NLO)   corrections to the production and decay of the top quark and the Higgs boson.
We find that QCD NLO  corrections can increase the total production cross sections by about 48.9\%  and  57.9\%  for the  $Htu$  and $Htc$ coupling induced processes at the LHC, respectively. After kinematic cuts are imposed on the decay products of the top quark and the Higgs boson, the QCD NLO corrections are reduced to $11\%$  for the $Htu$ coupling induced process and almost vanish for the $Htc$ coupling  induced process. Moreover,  QCD NLO corrections reduce the dependence of the total cross sections on the renormalization and factorization scales.
We also discuss  signals of the $tH$ associated production with the decay mode $t \rightarrow bl^{+}E\!\!\!\!\slash_{T},~H \rightarrow b\bar{b}$ and $t\bar{t}$ production with the decay mode $\bar{t} \rightarrow H\bar{q},~t\rightarrow bl^{+}E\!\!\!\!\slash_{T}$.
Our results show that, in  some parameter regions, the LHC may observe the above signals at the $5\sigma$ level.
Otherwise,  the upper limits on the  FCNC $Htq$  couplings can be set.
\end{abstract}

\maketitle

\section{INTRODUCTION}\label{s1}
Recently the ATLAS and CMS collaborations at the Large Hadron Collider (LHC)  have discovered  a new particle with a mass of about 125 GeV~\cite{:2012gu,:2012gk}. In the near future, the most important task is to study the intrinsic properties of this new particle, such as the couplings and spin, which will determine whether it is the Standard Model (SM) Higgs boson, and lead to deeper understanding of electroweak (EW) symmetry breaking mechanism.

It is attractive to investigate  the anomalous couplings of a generic Higgs boson, such as the  anomalous couplings with quarks via flavor-changing neutral-current (FCNC). In the SM, FCNC is absent at tree level, and is suppressed at one-loop level by the Glashow-Iliopoulos-Maiani mechanism~\cite{Glashow:1970gm}.  The anomalous couplings, if exist, are strong evidence of New Physics (NP). In Ref.~\cite{Blankenburg:2012ex}, indirect constraints on FCNC couplings from low-energy experiments are studied and used to analyze the process of the Higgs boson decaying to light quarks or leptons. But the FCNC couplings between the Higgs boson  and the top quark are not discussed there.

The top quark mass is close to the EW symmetry breaking scale. Thus it is an appropriate probe for the EW symmetry breaking mechanism and NP. Any deviation from SM prediction for precise observables  involving top quarks  exhibits  hints of NP. The production of a single top quark associated with a gluon jet or a vector boson via FCNC couplings has already been investigated at the leading order (LO)~\cite{AguilarSaavedra:2004wm,delAguila:1999ac,AguilarSaavedra:2000db,AguilarSaavedra:2000aj} and at the next-to-leading order (NLO)~\cite{Liu:2005dp,Zhang:2008yn,Gao:2009rf,Zhang:2011gh,Li:2011ek}, respectively. In this paper, we will discuss the constraints on the FCNC $Htq$ couplings from the signal of the Higgs boson and the top quark associated production with  QCD NLO accuracy  at the LHC.

In some NP models, the FCNC couplings of  the Higgs boson and the top quark can be generated at tree level, or enhanced to observable levels through radiative corrections~\cite{AguilarSaavedra:2004wm,Yang:2004af,delAguila:1999ac,AguilarSaavedra:2000db,AguilarSaavedra:2000aj,Liu:2005dp,Zhang:2008yn,Gao:2009rf,Zhang:2011gh,Li:2011ek}, such as the two Higgs doublet models III~\cite{Bejar:2003em,Cao:2003vf}, the minimal supersymmetric models (MSSM)~\cite{Guasch:1999jp,Bejar:2000ub,Yang:1993rb,deDivitiis:1997sh,DiazCruz:2001gf,Cao:2006xb}, the topcolor-assisted technicolor model~\cite{Cao:2003vf}, the exotic quarks models and the left-right symmetric models~\cite{Gaitan:2006eh,Arhrib:2006pm}. Since we do not know which type of NP will be responsible for the future deviation, it is better to study the FCNC processes with a model independent method. In general, the interactions between  the Higgs boson, the top quark and  a light quark $q ~(= u,c)$ can be expressed as~\cite{AguilarSaavedra:2004wm}
\begin{equation}\label{lag}
 \mathcal {L} = -\frac{g}{2 \sqrt 2} \sum_{q = u,c} g_{qt} \, \bar q (g_{qt}^v + g_{qt}^a \gamma_5) t H
+ \mathrm{H.c.},
\end{equation}
where $g_{qt} ~(q=u,c)$ defining the strength of the coupling  is  a real coefficient, while $g_{qt}^v$, $g_{qt}^a$ are complex numbers,
normalized to $|g_{qt}^v|^2+|g_{qt}^a|^2=1$.
The constraints on the above couplings have been set through indirect low-energy processes in Refs.~\cite{Fernandez:2009vr,Aranda:2009cd,Larios:2004mx}.
It is interesting to study how to set the direct constraints on  the above couplings from the signals of the top quark and the Higgs boson associated production and top pair production with top quark rare decay at the hadron colliders. These processes have been studied at the LO in Ref.~\cite{AguilarSaavedra:2004wm}. However, the LO total cross sections at the hadron collider suffer from large uncertainties due to the arbitrary choices of the renormalization and factorization scales. Besides, at the NLO level the additional radiation makes $b$ jets, which are the products of the Higgs boson and the top quark, softer and the mass distribution of the reconstructed particles broader, which will affect events selection when kinematic cuts are imposed. Thus, it is  necessary to perform complete QCD NLO  calculations of these processes  at the LHC, including production and decay.

The  process $gb \rightarrow tH^{-} $, which has the similar scattering amplitude to the process we study in this paper, has been discussed  in the two Higgs double models and supersymmetry, including the QCD NLO or supersymmetry QCD corrections~\cite{Plehn:2002vy,Weydert:2009vr,Zhu:2001nt,Klasen:2012wq}. After considering the difference from the couplings and parton distribution functions (PDFs), our numerical results for the $tH$ production are consistent with the their results in the range of Monte Carlo integration error. However, in the process we study, the Higgs boson is neutral and has a mass around 125 GeV, usually  much less than the mass of charged Higgs boson, which leads to significantly different decay modes and signals at the hadron colliders.

The arrangement of this paper is as follows. In Sec.~\ref{s2}, we present the LO results  for the Higgs boson and the top quark associated production induced by FCNC $Htq$ couplings. In Sec.~\ref{s3}, we describe the detailed calculations of the NLO results, including the virtual and real corrections. Then  in Sec.~\ref{s4}, we investigate numerical results, in which we discuss the scale uncertainties and give some important kinematic distributions. In Sec.\ref{s5}, we discuss the signals of $tH$ associated production with the decay mode $t\rightarrow bl^+ \nu_e$ and $H \rightarrow b\bar{b}$, and  $t\bar{t}$ production with rare decay mode $\bar{t}\rightarrow Hq\rightarrow b\bar{b}q$. Then we analysis the discovery potential with QCD NLO accuracy at the LHC with $\sqrt{S}=14$ TeV.  Section~\ref{s6} is a brief  conclusion.

\section{LEADING ORDER RESULTS}\label{s2}
\begin{figure}[h]
  \includegraphics[width=0.5\linewidth]{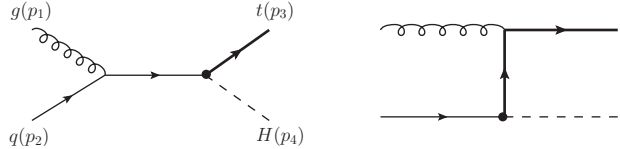}\\
  \caption{The LO Feynman diagrams for $tH$ associated production via FCNC $Htq$ couplings.}\label{eps:tree}
\end{figure}

At the hadron colliders, there is only one subprocess that contributes to the $tH$ associated production at the LO via FCNC  $Htq$ couplings:
\begin{equation}\label{process}
g ~ q \quad \rightarrow \quad t ~ H,
\end{equation}
where $q$ is either $u$ or $c$ quark. The Feynman diagrams are shown in Fig.~\ref{eps:tree}.

After summing (averaging) over the spins and colors of the final-(initial-)state particles, the explicit expression of the  squared amplitude at the LO is
\begin{eqnarray}\label{eq1}
\overline{|\mathcal {M}^B|^2}_{qg}(s,t)&=& \frac{ \pi ^2 \alpha  \alpha _s g_{qt}^2}{3\sin\theta_W^2 s\left(t-m_t^2\right)^2}
\big(m_t^6-(2m_H^2+2s+t)m_t^4+(2m_H^4+(s+t)^2)m_t^2-\nonumber \\
\nonumber\\
&&2t m_H^4+2t(s+t)m_H^2-t(s+t)^2\big),
\end{eqnarray}
where $m_t$ and $m_H$ are the top quark mass and the Higgs boson mass, respectively. The Mandelstam variables $s$, $t$, and $u$ are defined as
\begin{eqnarray}\label{eq2}
s = (p_1+p_2)^2,\  \ t = (p_1-p_3)^2,\  \ u = (p_1-p_4)^2.
\end{eqnarray}

The LO total cross section at hadron colliders is given by convoluting the partonic cross section with the PDFs $G_{i/P}$ in the proton,
\begin{equation}
\sigma^B=\int dx_1
dx_2\left[G_{g/P_1}(x_1,\mu_f)G_{q/P_2}(x_2,\mu_f) + (x_1 \leftrightarrow x_2)\right]\hat
\sigma^B_{qg},
\end{equation}
where $\mu_f$ is the factorization scale, and  $\hat \sigma^{B}_{qg}= (1/2 s)\int \overline{|\mathcal {M}^B|^2}_{qg} dPS^{(2)}$ is Born level partonic cross section.

\section{THE NEXT-TO-LEADING ORDER CALCULATIONS}\label{s3}
\begin{figure}[h]
  \includegraphics[width=1.0\linewidth]{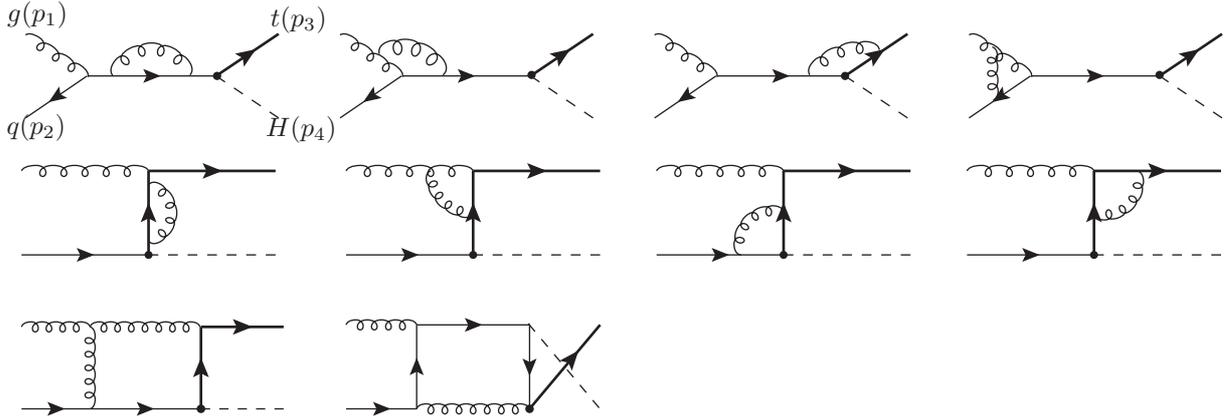}\\
  \caption{One-loop virtual Feynman diagrams for the $tH$ associated production via FCNC $Htq$ couplings.}\label{eps:virtual}
\end{figure}

In this section, we present QCD NLO  corrections of $tH$ associated production  using dimensional regularization scheme with naive $\gamma_5$ prescription in $n = 4- 2 \epsilon$ dimensions to regularize the ultraviolet (UV) and infrared (IR) divergences. The NLO corrections contain the virtual gluons effects and the real radiation of a gluon or a massless quark. The corresponding Feynman diagrams are shown in Fig.~\ref{eps:virtual} and~\ref{eps:real}, respectively. We use two cutoff phase space slicing method~\cite{Harris:2001sx} in the real corrections to separate the IR divergences.

\subsection{Virtual corrections}
The virtual corrections come from the interference of the one-loop amplitude with the Born amplitude:
\begin{eqnarray}\label{sigmaV}
\hat \sigma^V = \frac{1}{2s} \int{d}PS^{(2)} 2Re( \overline{\mathcal {M}^V \cdot \mathcal {M}^{B\ast} } ).
\end{eqnarray}
We introduce counterterms to absorb UV divergences. For the external fields, we fix all the renormalization constants using the on shell renormalization scheme:
\begin{eqnarray}\label{deltaZ}
\delta Z_2^{(g)}&=&-\frac{\alpha_s}{2\pi}C_{\epsilon}\left(
\frac{n_f}{3}-\frac{5}{2}\right)\left(\frac{1}{\epsilon_{UV}}-
\frac{1}{\epsilon_{IR}}\right)-\frac{\alpha_s}{6\pi}C_{\epsilon}
\frac{1}{\epsilon_{UV}}, \nonumber \\
\delta Z_2^{(q)}&=&-\frac{\alpha_s}{3\pi}C_{\epsilon}\left(\frac{1}
{\epsilon_{UV}}-\frac{1}{\epsilon_{IR}}\right), \nonumber \\
\delta Z_2^{(t)}&=&-\frac{\alpha_s}{3\pi}C_{\epsilon}\left(\frac{1}
{\epsilon_{UV}}+\frac{2}{\epsilon_{IR}}+4\right), \nonumber \\
\frac{\delta m_t}{m_t}&=&-\frac{\alpha_s}{3\pi}C_{\epsilon}\left(\frac
{3}{\epsilon_{UV}}+4\right),
\end{eqnarray}
where $C_{\epsilon}=\Gamma(1+\epsilon)\left[(4\pi\mu_r^2)/m_t^2\right]
^{\epsilon}$, $\mu_r$  is the renormalization scale and $n_f=5$.
For the counterterm of  FCNC couplings $\delta Z_{g_{qt}}$, we use the modified minimal subtraction (${\rm \overline{MS}}$) scheme~\cite{Zhang:2010bm}:
\begin{eqnarray}
\delta Z_{g_{qt}}&=& -\frac{\alpha_s}{4\pi ^2}\Gamma(1+\epsilon)(4\pi)^{\epsilon}\frac{1}{\epsilon _{UV}},
\end{eqnarray}
and the running of the FCNC coupling is given by
\begin{eqnarray}\label{running_gqt}
g_{qt}(\mu_r)=g_{qt}(\mu_0)\left(\frac{\alpha_s(\mu_r)}{\alpha_s(\mu_0)}\right)^{4/\beta_0}.
\end{eqnarray}
Here $\beta_0=11 -\frac{2}{3} n_f$ is the one-loop coefficients of the QCD $\beta$-function.

In the virtual corrections, the UV divergences are canceled by the counterterms.  To deal with IR divergences, we  adopt the traditional Passarino-Veltman reduction method to reduce the tensor integrals  to scalar integrals~\cite{Passarino:1978jh,Denner:1991kt}, of which the IR divergences can be obtained by the skill in Ref.~\cite{Ellis:2007qk}.  And the IR divergent parts are given by
\begin{eqnarray}
\mathcal {M}^V|_{IR} &=& \frac{\alpha_s}{2\pi} \frac{\Gamma(1-\epsilon)}{\Gamma(1-2\epsilon)}\left(\frac{4\pi\mu_r^2}{s}\right)^{\epsilon} \left(\frac{A_2^v}{\epsilon_{IR}^2}
+\frac{A_1^v}{\epsilon_{IR}}\right)\mathcal {M}^B,
\end{eqnarray}
where
\begin{eqnarray}
 A_2^v &=& -\frac{13}{3}, \nonumber \\
 A_1^v &=& 3\ln\frac{m_t^2-t}{m_{t}^2} -\frac{1}{3}\ln\frac{m_t^2-u}{m_{t}^2} -\frac{4}{3}\ln\frac{s}{m_{t}^2}
-\frac{43}{6}.
\end{eqnarray}

\subsection{Real corrections}
\begin{figure}[h]
  \includegraphics[width=1.0\linewidth]{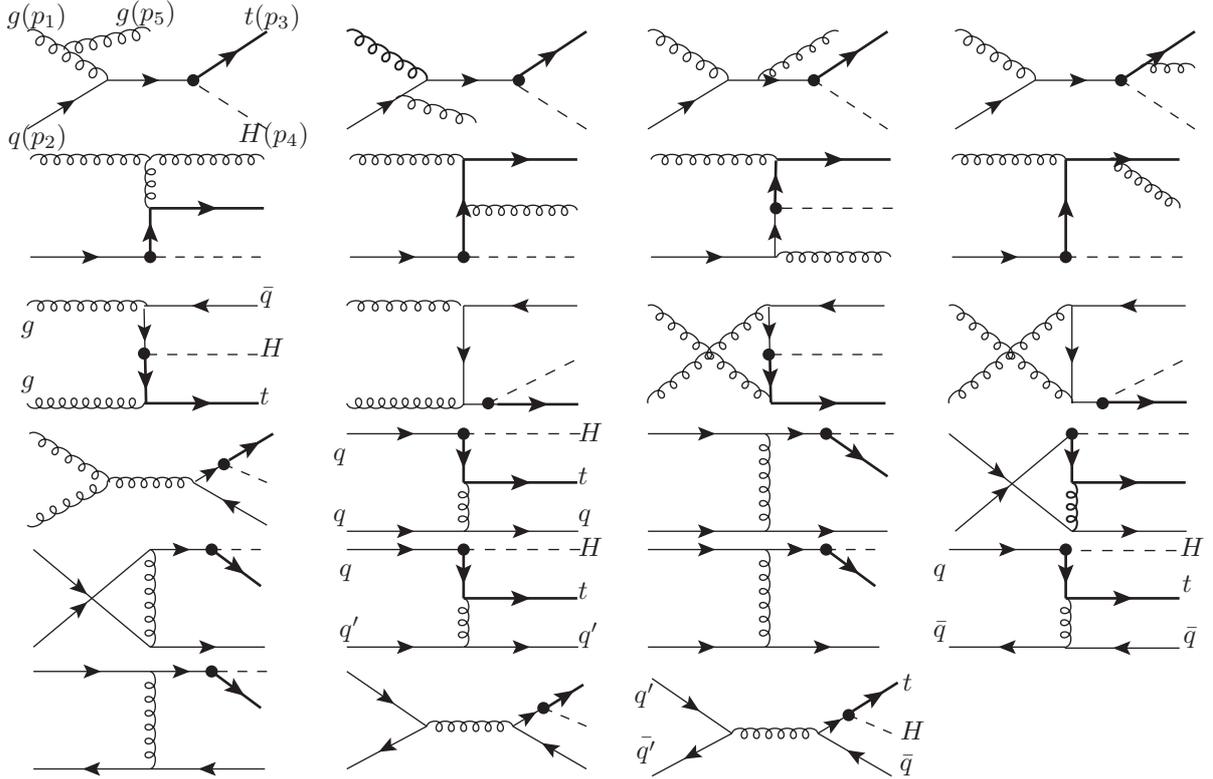}\\
  \caption{Feynman diagrams for the real corrections to the $tH$ associated production via FCNC $Htq$  couplings.}\label{eps:real}
\end{figure}

The real corrections contain the radiations of an additional gluon, and massless (anti)quark. The relevant Feynman diagrams are shown in Fig.~\ref{eps:real}.
The partonic cross section can be written as
\begin{eqnarray}\label{delta Z_gqt}
\hat \sigma^R=\frac{1}{2s}\int{d}PS^{(3)}\overline{|\mathcal {M}^R|^2}.
\end{eqnarray}

We use two cutoff phase space slicing method~\cite{Harris:2001sx}, which introduces two small cutoffs $\delta_s$ and $\delta_c$ to divide the three-body phase space into three regions. First,  the phase space is separated into  two regions according to whether  or not the  energy of the additional gluon satisfies the soft criterion $E_5 \leq \delta_s\sqrt{s}/2$ in the partonic center-of-mass frame (CMF).  And the partonic cross section can be divided as
\begin{equation}
\hat{\sigma }^R = \hat{\sigma }^{\text{H}}+\hat{\sigma }^S,
\end{equation}
where $\hat{\sigma }^{\text{H}}$ and $\hat{\sigma }^S$ are the contributions from the hard and the soft regions, respectively.
Furthermore,  the collinear cutoff $\delta_c$ is applied to separate the hard region into two  regions according to whether the collinear condition $-\delta_c s< t_{i5} < 0$ is satisfied or not, where $t_{i5} = (p_i - p_5)^2$ with $i=1,2$.  The corresponding cross section is splited into
\begin{equation}
\hat{\sigma }^{\text{H}} = \hat{\sigma }^{\overline{\text{HC}}}+\hat{\sigma }^{\text{HC}}.
\end{equation}
where  the contributions from hard-collinear regions $\hat{\sigma }^{\text{HC}}$ contain the collinear divergences, and  the hard-noncollinear part $\hat{\sigma }^{\overline{\text{HC}}}$ is free of IR singularities and  can be calculated numerically.

\subsubsection{Real gluon emission}
In the limit that the energy of the emitted gluon becomes small, i.e., $E_5 \leq \delta_s\sqrt{s}/2$, the amplitude squared
 $\overline{|\mathcal {M}(qg \to t H +g)|^2}$  can be factorized into the Born amplitude squared
 and the eikonal factor $\Phi_{\text{eik}}$:
\begin{equation}
\overline{|\mathcal {M}(qg \to t H +g)|^2}
\stackrel{\text{soft}}{\longrightarrow}
(4\pi\alpha_s\mu_r^{2\epsilon}) \overline{|\mathcal {M}^B|^2}
\Phi_{\text{eik}},
\end{equation}
where the eikonal factor can be expressed as
\begin{eqnarray}
\Phi_{\text{eik}}&=&\frac{C_A}{2}\frac{s}{(p_{1}\cdot p_{5})(p_{2}\cdot p_{5})}-\frac{1}{2C_A}\frac{m_t^2-u}{(p_{2}\cdot p_{5})(p_{3}\cdot p_{5})} \nonumber\\
&&+\frac{C_A}{2}\frac{m_t^2-t}{(p_{2}\cdot p_{5})(p_{3}\cdot
p_{5})}-C_F\frac{m_t^2}{(p_{3}\cdot p_{5})^2},
\end{eqnarray}
with $C_A=3, C_F=\frac{4}{3}$. The three-body phase space in the soft region can be factorized into
\begin{equation}
dPS^{(3)}(qg \to tH
+g)\stackrel{\text{soft}}{\longrightarrow} dPS^{(2)}(qg \to
tH)dS,
\end{equation}
where $dS$ is the integration over the phase space of the soft gluon, given by
\begin{equation}
dS = \frac{1}{2(2\pi)^{3-2\epsilon}} \int^{\delta_s \sqrt{s}/2}_0
dE_5E_5^{1-2\epsilon}d\Omega_{2-2\epsilon}.
\end{equation}
Hence, the parton level cross section in the soft region can be expressed as
\begin{equation}\label{soft1}
 \hat{\sigma}^S =
(4\pi\alpha_s\mu^{2\epsilon}_r)
\hat\sigma^{B}\int  \Phi_{\text{eik}} dS.
\end{equation}
After integration over
the soft phase space, Eq.~(\ref{soft1}) becomes
\begin{equation}
\hat{\sigma}^S = \hat{\sigma}^B
\left[\frac{\alpha_s}{2\pi}\frac{\Gamma(1-\epsilon)}{\Gamma(1-2\epsilon)
}\left(\frac{4\pi\mu^2_r}{s}\right)^{\epsilon}\right]\left(\frac{A^s_2}
{\epsilon_{IR}^2}+\frac{A^s_1}{\epsilon_{IR}}+A^s_0\right),
\end{equation}
with
\begin{eqnarray}
&& A_2^s=\frac{13}{3},
\nonumber \\
&& A_1^s= -2A_2^s\ln\delta_s  +\frac{1}{3}
\ln\frac{m_t^2-u}{m_t^2} -3
\ln\frac{m_t^2-t}{m_t^2}
+\frac{4}{3}\ln\frac{s}{m_t^2}+\frac{4}{3},
\nonumber \\
&& A_0^s=-2A_2^s\ln^2\delta_s-2A_1^s \ln\delta_s + \frac{4}{3\beta}\ln \frac{1+\beta}{1-\beta}
+\frac{3}{2} B_{-} - \frac{1}{6} B_{+}.
\end{eqnarray}
The $B_{\pm}$ are defined as
\begin{eqnarray}
&& B_{\pm} =\ln^2\frac{1-\beta}{1\pm\beta \cos\theta}
-\frac{1}{2}\ln^2\frac{1 +\beta}{1 -\beta} +2{\rm
Li}_2\left[\frac{\mp\beta (\cos\theta \pm1)}{1-\beta }\right]
\nonumber \\
&& \hspace{1.0cm} -2{\rm
Li}_2\left[\frac{\pm\beta (\cos\theta \mp1)}{1\pm\beta \cos\theta }\right],
\end{eqnarray}
where $\beta=\sqrt{1-4 m_t^2 s/(m_t^2-m_H^2+s)^2}$ and $\cos\theta = (t-u)/\sqrt{(m_t^2-m_h^2+s)^2-4m_t^2s}$.

In the hard-collinear region, collinear singularities arise when  the emitted hard gluon is collinear to the incoming massless partons. As the conclusion of the factorization theorem~\cite{Collins:1985ue,Bodwin:1984hc}, the amplitude squared $\overline{|\mathcal {M}(qg \to t H +g)|^2}$  can be factorized into the product of the Born amplitude squared
and the Altarelli-Parisi splitting function; that is
 \begin{equation}
\overline{|\mathcal {M}(qg \rightarrow tH +
g)|^2}\stackrel{\text{coll}}
{\longrightarrow}(4\pi\alpha_s\mu^{2\epsilon}_r)\overline{|\mathcal {M}^B|^2}\left(\frac{-2P_{gg}(z,\epsilon)}{zt_{15}}
+\frac{-2P_{qq}(z,\epsilon)}{zt_{25}}\right),
\end{equation}
where $z$ denotes the fraction of the momentum of the incoming parton carried by $q(g)$.
The unregulated Altarelli-Parisi splitting functions are written explicitly as~\cite{Harris:2001sx}
 \begin{eqnarray}
P_{qq}(z,\epsilon) &=& C_{F}\Big(\frac{1+z^{2}}{1-z}-\epsilon(1-z)\Big), \nonumber \\
P_{gg}(z,\epsilon) &=& 2C_A\Big(\frac{z}{1-z}+\frac{1-z}{z}+z(1-z)\Big).
\end{eqnarray}
Then we factorize the three-body phase space in the collinear limit $-\delta_c s < t_{i5} < 0$ as
\begin{equation}
dPS^{(3)}(qg \rightarrow tH +
g)\stackrel{\text{coll}}{\longrightarrow} dPS^{(2)}(qg
\rightarrow tH; s^{\prime} = zs)
\frac{(4\pi)^{\epsilon}}{16\pi^2\Gamma(1-\epsilon)}dzdt_{i5}[-(1-z)t_{i5}]^{-\epsilon}.
\end{equation}
Thus, after convoluting with the PDFs, the three-body cross section
in the hard-collinear region can be written as~\cite{Harris:2001sx}
\begin{eqnarray}
\sigma^{HC} & = & \int dx_1dx_2 ~\hat{\sigma}^B_{qg}
\left[\frac{\alpha_s}{2\pi}\frac{\Gamma(1-\epsilon)}{\Gamma(1-2\epsilon)}
\left(\frac{4\pi\mu^2_r}{s}\right)^{\epsilon}\right]\left(-\frac{1}{\epsilon}\right)
\delta_c^{-\epsilon}\left[P_{qq}(z,\epsilon)G_{q/p}(x_1/z)G_{g/p}(x_2)
\right.\nonumber\\&& \left.+
P_{gg}(z,\epsilon)G_{g/p}(x_2/z)G_{q/p}(x_1) +
(x_1 \leftrightarrow x_2)\right]
\frac{dz}{z}\left(\frac{1-z}{z}\right)^{-\epsilon},
\end{eqnarray}
where $G_{q(g)/p}(x)$ is the bare PDF.

\subsubsection{Massless (anti)quark emission}
In addition to the real gluon emission, an additional massless
$q(\bar q)$ in the final state should be taken into consideration at $\mathcal {O}(\alpha_s)$ of the perturbative expansion. Since the contributions from real massless $q(\bar q)$ emission contain initial-state collinear singularities, we need to use the two cutoff phase space slicing method~\cite{Harris:2001sx} to isolate these collinear divergences.  The
cross section for the process with an additional massless
$q(\bar{q})$ emission, including the noncollinear part and collinear part, can be expressed as
\begin{eqnarray}
\label{sigma:nc} \sigma^{add} & = & \int dx_1dx_2\sum_{(\alpha=q,\bar{q},q')} \Big\{
\hat{\sigma}^{\overline{C}}(q \alpha \rightarrow tH + q(\bar{q}))G_{q/p}(x_1)G_{\alpha/p}(x_2)+
\hat{\sigma}^B_{qg} \left[\frac{\alpha_s}{2\pi}\frac{\Gamma(1-\epsilon)}{\Gamma(1-2\epsilon)} \left(\frac{4\pi\mu^2_r}{s}\right)^{\epsilon}\right] \nonumber\\
&& \left(-\frac{1}{\epsilon}\right) \delta_c^{-\epsilon}
P_{g\alpha}(z,\epsilon)G_{q/p}(x_1)G_{\alpha/p}(x_2/z)\frac{dz}{z}\left(\frac{1-z}{z}\right)^{-\epsilon}+ (x_1\leftrightarrow x_2)\Big\}+  \nonumber\\
&& \int dx_1dx_2 ~\Big\{\hat{\sigma}^{\overline{C}} (gg \rightarrow tH +\bar{q})G_{g/p}(x_1)G_{g/p}(x_2)+
\hat{\sigma}^B_{qg} \left[\frac{\alpha_s}{2\pi}\frac{\Gamma(1-\epsilon)}{\Gamma(1-2\epsilon)} \left(\frac{4\pi\mu^2_r}{s}\right)^{\epsilon}\right] \nonumber\\
&& \left(-\frac{1}{\epsilon}\right) \delta_c^{-\epsilon} P_{qg}(z,\epsilon)G_{g/p}(x_1/z)G_{g/p}(x_2) \frac{dz}{z}\left(\frac{1-z}{z}\right)^{-\epsilon}+ (x_1 \leftrightarrow x_2)\Big\},
\end{eqnarray}
where
\begin{eqnarray}
P_{qg}(z,\epsilon) & = & P_{\bar{q}g}(z) =
\frac{1}{2}\left[z^2+(1-z)^2\right]-z(1-z)\epsilon,\nonumber\\
P_{gq}(z,\epsilon) & = & P_{g\bar{q}}(z) = C_F\left[\frac{z}{1+(1-z)^2}-z\epsilon\right].
\end{eqnarray}
The $\hat{\sigma}^{\overline{C}}$ is the noncollinear cross sections for the processes of the $q\bar{q}(q,q')$, $gg$ and $q'\bar{q'}$ initial states:
\begin{eqnarray}
\hat{\sigma}^{\overline{C}} &=& \frac{1}{2s}\int\Big\{\sum_{(\alpha=q,\bar{q},q')}
|\mathcal {M}(q\alpha\longrightarrow tH+\alpha)|^2+ |\mathcal {M}(gg \longrightarrow tH+\bar{q})|^2 + \nonumber \\
&& |\mathcal {M}(q'\bar{q'}\longrightarrow tH+\bar{q})|^2\Big\}d PS^{(3)}_{~\overline{C}},
\end{eqnarray}
in which $d PS^{(3)}_{~\overline{C}}$ is the three-body phase space in the
noncollinear region.

\subsubsection{Mass factorization}
There are still  collinear
divergences in the partonic cross sections after adding the renormalized virtual corrections and the real
corrections. The remaining divergences can be factorized into a redefinition of the PDFs. In the
$\overline{\text{MS}}$ convention the scale-dependent PDF
$G_{\alpha/p}(x,\mu_f)$ can be written as ~\cite{Harris:2001sx}
\begin{eqnarray}
\label{modifiedPDF} G_{\alpha/p}(x,\mu_f) & = & G_{\alpha/p}(x) +
\sum_{\beta}\left(-\frac{1}{\epsilon}\right)\left[
\frac{\alpha_s}{2\pi}\frac{\Gamma(1-\epsilon)}{\Gamma(1-2\epsilon)}
 \left(\frac{4\pi\mu^2_r}{\mu_f^2}\right)^{\epsilon}\right]\nonumber\\
&& \times \int_x^1 \frac{dz}{z} P_{\alpha\beta}(z) G_{\beta/p}(x/z).
\end{eqnarray}
The Altarelli-Parisi splitting function is defined by
\begin{eqnarray}
P_{\alpha\beta}(y,\epsilon) = P_{\alpha\beta}(y) +\epsilon P'_{\alpha\beta}(y).
\end{eqnarray}
The resulting $\mathcal{O} (\alpha_s)$ expression for the initial-state collinear contribution
can be written in the following form:
\begin{eqnarray}
 \sigma^{coll} &=&  \int dx_1 dx_2 \hat{\sigma}^B_{qg}\bigg[\frac{\alpha_s}{2\pi}
\frac{\Gamma(1-\epsilon)} {\Gamma(1-2\epsilon)}
\bigg(\frac{4\pi\mu^2_r}{s}\bigg)^\epsilon \bigg] \nonumber\\
&&\bigg\{\tilde{G}_{q/p}(x_1,\mu_f) G_{g/p}(x_2,\mu_f) +
G_{q/p}(x_1,\mu_f) \tilde{G}_{g/p}(x_2,\mu_f)+ \nonumber
\\ &&
\sum_{\alpha=q,g}\bigg[\frac{A_1^{sc}(\alpha\rightarrow
\alpha g)}{\epsilon} +A_0^{sc}(\alpha\rightarrow \alpha
g)\bigg]G_{q/p}(x_1,\mu_f) G_{g/p}(x_2,\mu_f)+ \nonumber
\\ &&
(x_1\leftrightarrow x_2)\bigg\} ,\label{11}
\end{eqnarray}
where
\begin{eqnarray}
A_1^{sc}(q\rightarrow qg)&=&C_F(2\ln\delta_s +\frac{3}{2}), \nonumber\\
A_1^{sc}(g\rightarrow gg)&=&2C_A\ln\delta_s + \frac{11C_A-2n_f}{6}, \nonumber\\
A_0^{sc}&=&A_1^{sc}\ln\left(\frac{s}{\mu_f^2}\right),
\end{eqnarray}
and
\begin{eqnarray}
\tilde{G}_{\alpha/p}(x,\mu_f)&=&\sum_{\beta}\int_x^{1-
\delta_s\delta_{\alpha\beta}} \frac{dy}{y}
G_{\beta/p}\left(\frac{x}{y},\mu_f\right)\tilde{P}_{\alpha\beta}(y),
\end{eqnarray}
with
\begin{eqnarray}
\tilde{P}_{\alpha\beta}(y)=P_{\alpha\beta}(y) \ln\left(\delta_c
\frac{1-y}{y} \frac{s}{\mu_f^2}\right) -P_{\alpha\beta}'(y).
\end{eqnarray}

Finally, we combine all the results above  to give the NLO total cross section for the $pp\rightarrow tH$ process:
\begin{eqnarray}
&& \sigma^{NLO}= \int dx_1dx_2
\left\{\left[G_{q/p}(x_1,\mu_f)G_{g/p}(x_2,\mu_f)+ x_1\leftrightarrow
x_2\right](\hat{\sigma}^{B} + \hat{\sigma}^{V}+
\hat{\sigma}^{S} +\hat{\sigma}^{\overline{HC}})\right\}
+\sigma^{coll}\ \nonumber
\\ && \hspace{1.0cm} +\sum_{(\alpha,\beta)}\int dx_1dx_2
\left[G_{\alpha/p}(x_1,\mu_f) G_{\beta/p}(x_2,\mu_f)
+(x_1\leftrightarrow x_2)\right]
\hat{\sigma}^{\overline{C}}(\alpha\beta\rightarrow
tH + X),
\end{eqnarray}
where ($\alpha$, $\beta$) stand for the $q\bar{q}(q,q')$ , $gg$ and $q'\bar{q'}$ initial states.
Note that there  contain no singularities any more, since
$A_2^v +A_2^s =0$ and $A_1^v +A_1^s +A_1^{sc}(q\rightarrow qg)
+A_1^{sc}(g\rightarrow gg) =0$.

\section{Production Cross Section}\label{s4}
In this section, we give the numerical results for the total and differential cross sections for the Higgs boson and the top quark associated  production via the FCNC $Htq$ couplings.

\subsection{Experiment Constraints}
Before proceeding, we discuss the choice of the FCNC $Htq$ couplings, which is constrained by low-energy data on flavor-mixing process.
For example, in the precision measurement of the magnetic dipole moments of the proton and the neutron,  the contribution of $g_{ut}$ vertex should be less than the experimental uncertainty~\cite{Fernandez:2009vr,Aranda:2009cd}.  The result reveals that $Im(g_{qt}^a)$ is more strongly suppressed than  $Re(g_{ut}^v)$. Therefore the FCNC $Htq$ coupling can be described  by only one real parameter, i.e.,  the strength of the coupling $g_{qt}$, which is constrained as
\begin{eqnarray}
\left(\frac{g}{2\sqrt{2}}g_{ut}\right)^2<54\pi^2\frac{\Delta a_p^{Exp}}{x_t x_p g(x_t)},
\end{eqnarray}
where
\begin{eqnarray}
g(x) = \frac{3+x^2(x^2-4)+2\log(x^2)}{2(x^2-1)^3},
\end{eqnarray}
with  $x_a=m_a/m_H$.

In $D^0$-$\bar{D^0}$ mixing, the $g_{ut}$ and $g_{ct}$ couplings can contribute to the mass difference $\Delta M_D$ through loop effects.  Experiment results impose limits on $g_{ut}$ and $g_{ct}$ as~\cite{Fernandez:2009vr,Aranda:2009cd}:
\begin{eqnarray}
|g_{ut}g_{ct}|<\frac{1.73\times10^{-2}}{\sqrt{f(x)-2g(x)}},
\end{eqnarray}
where
\begin{eqnarray}
&&f(x) = \frac{1}{2}\frac{1}{(1-x)^3}(1-x^2+2x \log x),\nonumber\\
&&g(x) = \frac{4}{(1-x)^3}(2(1-x)+(1+x)\log x),
\end{eqnarray}
with $x=m_H^2/m_t^2$.

\begin{table}[h]
\begin{center}
\begin{tabular}{cccccccc}
  \hline
  \hline
   &   couplings                    &&  upper Limits                       && Experiments                 \\
  \hline
   &  $g_{ut}$      \quad \quad     &&  $0.363   \sim 0.393 $              && magnetic dipole moments~\cite{Fernandez:2009vr}     \\
  \hline
   &  $g_{ut}g_{ct}$   \quad \quad  &&  $0.0194  \sim 0.0272$              && $D^0$-$\bar{D^0}$ mixing~\cite{Fernandez:2009vr}    \\
  \hline
   &  $g_{ct}$    \quad \quad       &&  $0.270   \sim 0.319 $              && $Z \rightarrow c\bar{c}$~\cite{Larios:2004mx}   \\
  \hline
  \hline
\end{tabular}
\end{center}
 \caption{The $95\%$ C.L. upper limits on the FCNC $Htq$ couplings obtained from the low-energy experiments, with Higgs boson mass in the range from $115$ GeV to $170$ GeV.}\label{t_limit}
\end{table}

The EW precision observables $\Gamma_{Z}, R_{c}, R_{b}, R_{l}, A_{c}$ and $A_{FB}^c$  also impose constraints on the couplings $g_{ct}$ through the radiative corrections to the effective $Zc\bar{c}$ vertex~\cite{Larios:2004mx}. In Table~\ref{t_limit}, we list the $95\%$ C.L. upper limits on the FCNC $Htq$ couplings obtained from the low-energy experiments.

\subsection{Cross sections at the LHC}
Now we discuss the numerical results of the Higgs boson and the top quark associated  production at the LHC. Unless specified otherwise, we choose $g_{ut}=0.2$ and $g_{ct}=0.2$, which are allowed by the low-energy experiments. We have checked that the imaginary part of these couplings do not contribute to the final results, so   $Im(g_{qt}^a)$  is neglected in  numerical calculations. Other SM input parameters are:
\begin{eqnarray}\label{sm_para}
 && m_t=173.1 \textrm{~GeV},    \quad  m_H=125 \textrm{~GeV}, \quad   m_W= 80.398 \textrm{~GeV}, \quad    m_b(m_b) = 4.2 \textrm{~GeV},\nonumber \\
 && \quad \quad \quad\alpha=1/127.921,\quad  \sin^2\theta_{W}=0.2312,  \quad  \alpha_s(m_Z)=0.118.
\end{eqnarray}
The CTEQ6L (CTEQ6M) PDF sets and the corresponding running QCD coupling constant $\alpha_s$  are used in the LO (NLO) calculations. The factorization and renormalization scales are set as $\mu_f=\mu_r=\mu_0$, and $\mu_0=m_t+m_H$. Moreover, as to the Yukawa couplings of
the bottom quark, we take the running mass $m_b(Q)$ evaluated by the NLO formula~\cite{Carena:1999py}
\begin{align}
&m_b(Q)=U_6(Q,m_t)U_5(m_t,m_b)m_b(m_b),
\end{align}
The evolution factor $U_f$ is given by
\begin{eqnarray}
U_f(Q_2,Q_1)=\bigg(\frac{\alpha_s(Q_2)}{\alpha_s(Q_1)}\bigg)^{d^{(n_f)}}
\bigg[1+\frac{\alpha_s(Q_1)-\alpha_s(Q_2)}{4\pi}J^{(n_f)}\bigg],
\end{eqnarray}
with
\begin{eqnarray}
d^{(n_f)}=\frac{12}{33-2n_f}, \hspace{1.0cm}
J^{(n_f)}=-\frac{8982-504n_f+40n_f^2}{3(33-2n_f)^2}.
\end{eqnarray}
\begin{figure}[h]
      \begin{center}
     \scalebox{0.32}{\includegraphics*{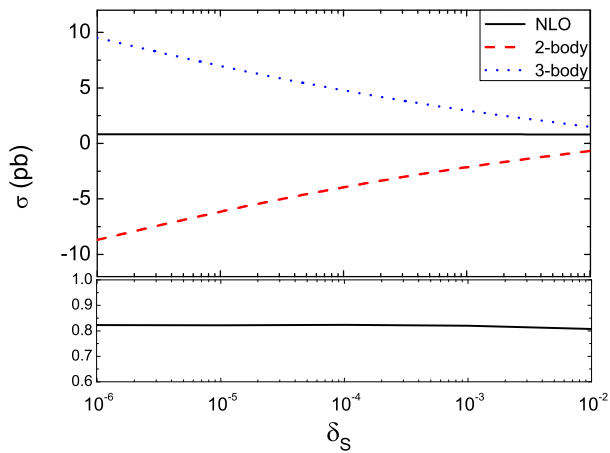}}
      \end{center}
    \caption{\label{udeltas}Dependence of the NLO total cross sections  at the LHC on the cutoff scale $\delta_{s}$ with  $\delta_{c}=\delta_{s}/50$.
     The label 2-body represents the  LO cross section, the one-loop virtual corrections, the soft and hard-collinear limits of the cross sections of three-body final states, while the label 3-body denotes the cross sections of hard-non-collinear part.
 }
\end{figure}

In Fig.~\ref{udeltas}, we show the dependence of the NLO cross sections on the cutoffs $\delta_s$ and $\delta_c$. In fact, the soft-collinear and the hard-noncollinear parts individually  strongly depend on the cutoffs. But the total cross section is independent on cutoffs after all pieces are added together. From Fig.~\ref{udeltas}, we can see that the change of $\sigma^{NLO}$ is very slow for $\delta_s$ in the range  from  $10^{-6}$ to $10^{-2}$, which indicates that it is reasonable to use the two cutoff phase space slicing method.

\begin{figure}[h]
        \begin{center}
     \scalebox{0.32}{\includegraphics*{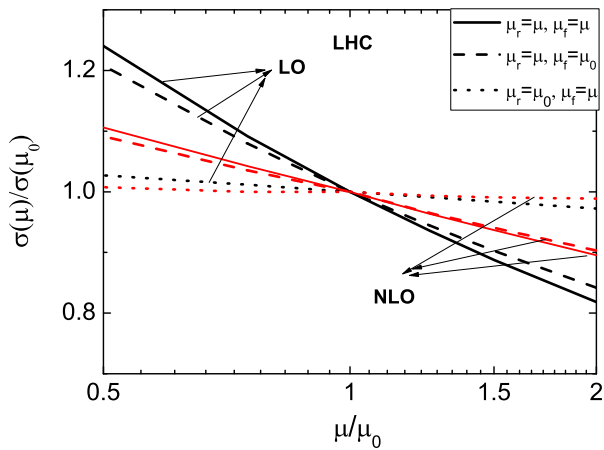}}
      \end{center}
  \caption{\label{scalein}Scale dependences of the cross sections for the $gu$ initial state subprocess  at the LHC.}
\end{figure}

In Fig.~\ref{scalein}, we give the scale dependences of the LO and NLO total cross sections. Explicitly, we consider three cases: $(1)$ factorization scale dependence ($\mu_r=\mu_0$ and $\mu_f = \mu$); $(2)$ renormalization  scale dependence  ($\mu_f=\mu_0$ and  $\mu_r = \mu$); $(3)$  total scale dependences ($\mu_f =\mu_r =\mu$). From Fig.~\ref{scalein}, we find that the NLO corrections significantly reduce the scale dependences  for all three cases, which makes the theoretical predictions more reliable.

\begin{figure}[h]
      \begin{center}
     \scalebox{0.32}{\includegraphics*{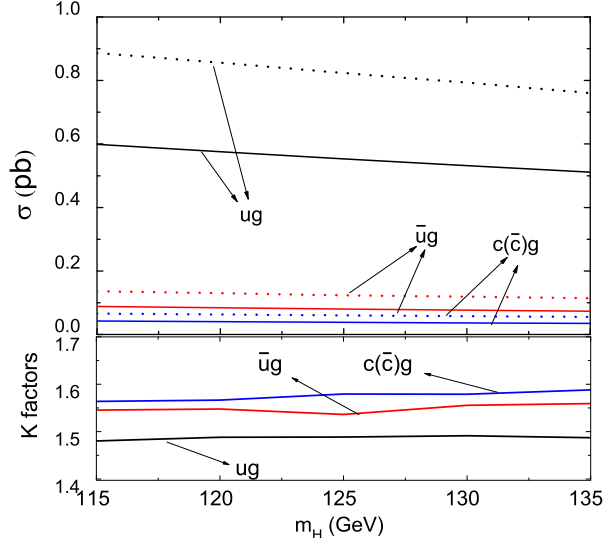}}
      \end{center}
 \caption{The LO and NLO total cross sections and K factors as functions of $m_H$. The solid lines and the dotted lines represent the LO and NLO cross sections, respectively.}\label{t_mass}
\end{figure}

Figure~\ref{t_mass} presents the  dependence of the total cross sections on $m_H$. It can be seen that the cross sections decrease by about $16\%$ as $m_H$  increases from $115$ GeV to $135$ GeV. In Fig.~\ref{t_mass}, the K factors, defined as $\sigma^{NLO}/\sigma^{LO}$, are also shown.  We can see that the  K factors  for  $ug \rightarrow tH$,  $\bar{u}g \rightarrow \bar{t}H$ and  $c(\bar{c})g \rightarrow tH$  processes are   $1.49$,  $1.54$ and $1.58$, respectively,  at the LHC for $m_H=125$ GeV, and they are not sensitive to the Higgs boson mass.

\begin{figure}[h]
\begin{minipage}[t]{0.45\linewidth}
\centering
  \includegraphics[width=1.1\linewidth]{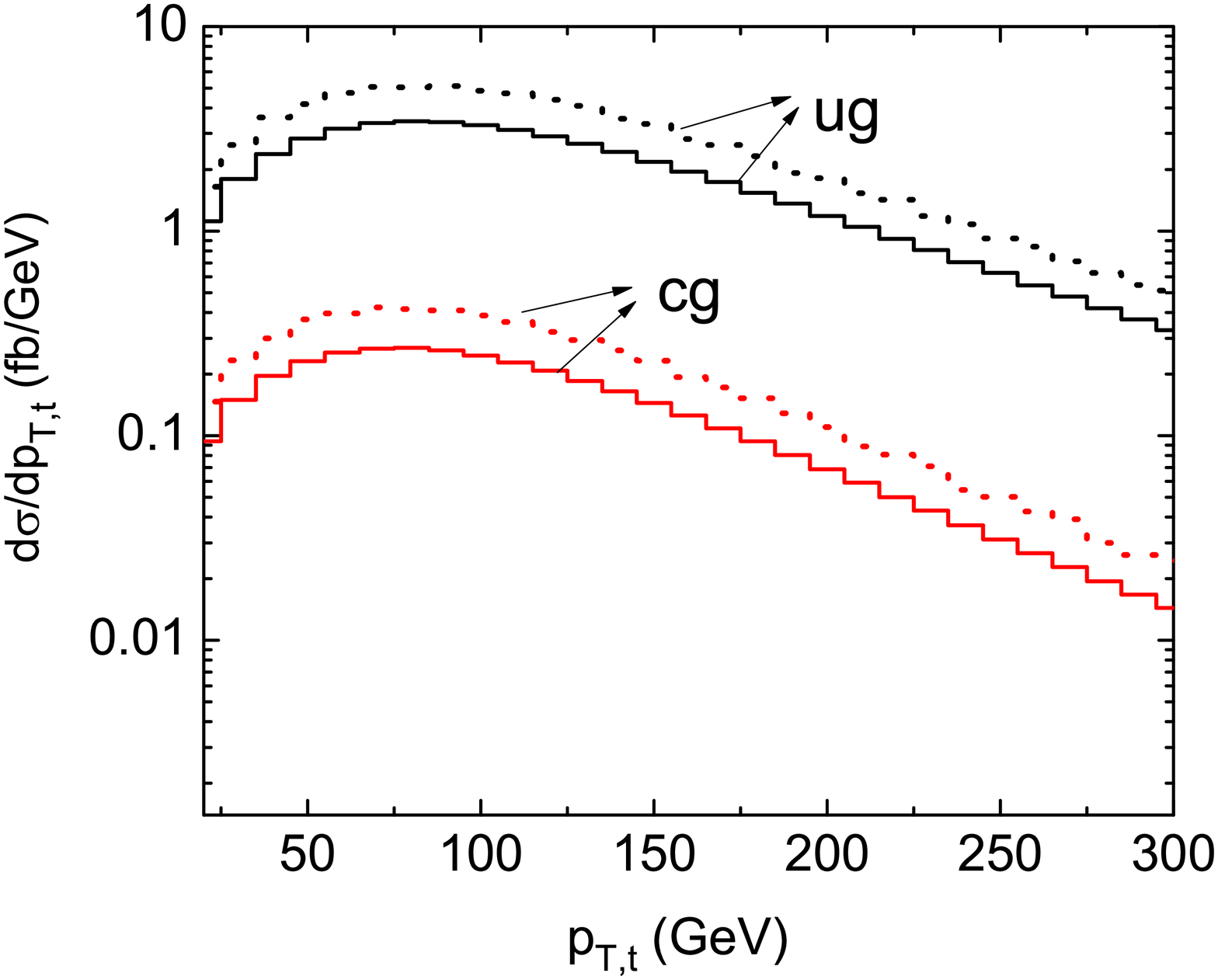}\\
\end{minipage}
\hfill
\begin{minipage}[t]{0.45\linewidth}
\centering
 \includegraphics[width=1.1\linewidth]{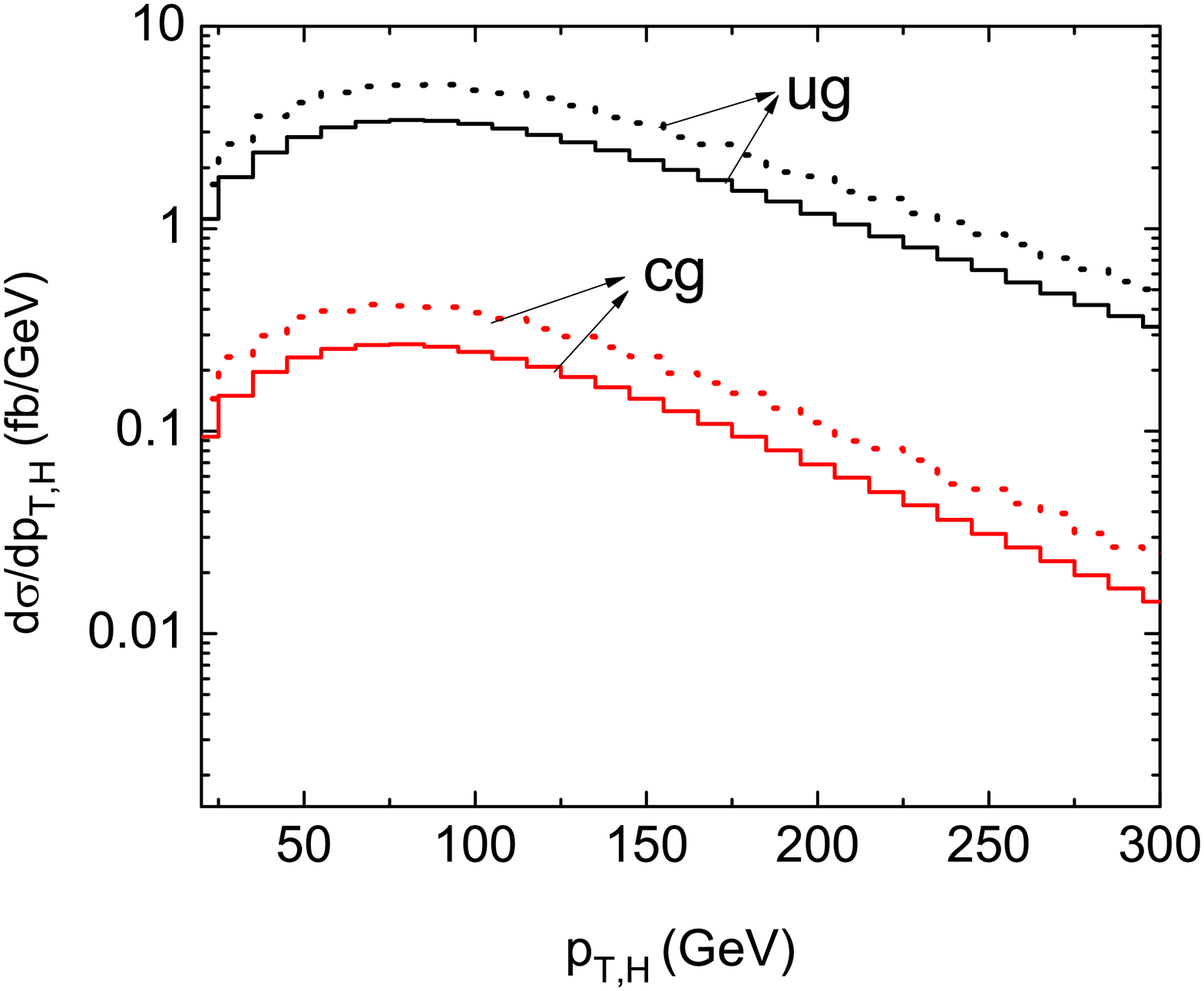}\\
\end{minipage}
\caption{\label{propt}The transverse momentum distributions of the top quark and the Higgs boson at the LHC. The solid line and the dotted line represent the LO and the NLO results, respectively.}
\end{figure}

\begin{figure}[h]
  \includegraphics[width=0.5\linewidth]{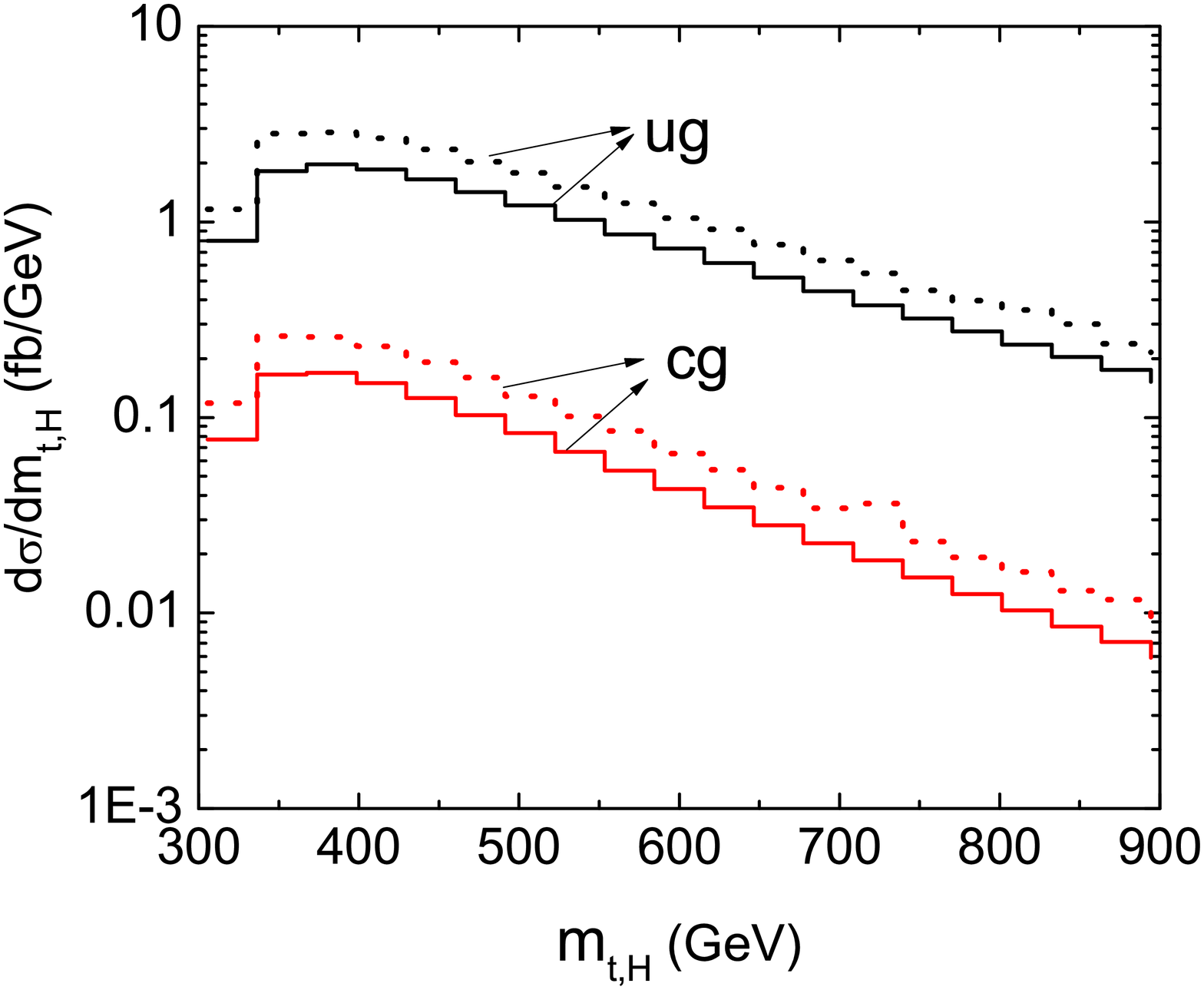}\\
  \caption{\label{proinm}Distributions of the invariant mass of the top quark and Higgs boson. The solid line and the dotted line represent the LO and the NLO results respectively.}
\end{figure}

Figure~\ref{propt} shows the differential cross sections as a function of the transverse momentum of the top quark and the Higgs boson. We find that they are very similar and the peak positions are around $80 ~{\rm GeV}$.
In  Fig.~\ref{proinm}, we present the distributions of  the invariant mass of the top quark and Higgs boson, and the peaks are around $360 ~{\rm GeV}$. From these figures, we can see that the NLO corrections significantly increase the LO results, but do not change the distribution shapes.

\section{Signal and discovery potentiality}\label{s5}
In this section, we investigate the signal and corresponding backgrounds in detail and present the discovery potential of the signal of the $tH$ associated production  at the LHC.
\subsection{NLO prediction on FCNC $tH$ associated production and decay}
\begin{figure}[h]
  \includegraphics[width=0.7\linewidth]{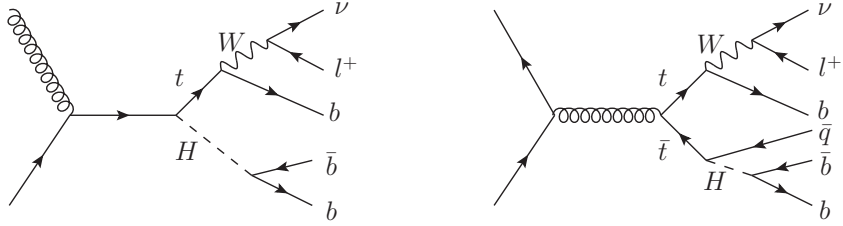}\\
  \caption{Representative Feynman diagrams for $tH$ production and $t\bar{t}\rightarrow tH\bar{q}$ production via FCNC $Htq$  couplings.}\label{f_signature}
\end{figure}

We have discussed the QCD NLO corrections to $tH$ associated production  in the last section.
In order to provide a complete QCD NLO prediction on the signal, we need to include the QCD NLO  corrections to the decay of the top quark and the Higgs boson. In this work, we concentrate on the top quark semileptonic decay and the Higgs boson decaying to $b\bar{b}$, as shown in Fig.~\ref{f_signature}.

The complete NLO cross section for $tH$ associated production and decay can be written as
\begin{eqnarray}\label{pro_dec_cro}
\sigma^{NLO}\left(pp\rightarrow t\left(\rightarrow \nu_{l}l^{+}b\right)+ H\left(\rightarrow b\bar{b}\right) \right) = \left(\sigma_0 + \alpha_s\alpha_1\right) \times \left(\frac{\delta\Gamma_0^H +\alpha_s \delta\Gamma_1^H}{\Gamma_0^H + \alpha_s\Gamma_1^H}\right) \left(\frac{\delta\Gamma_0^t +\alpha_s \delta\Gamma_1^t}{\Gamma_0^t + \alpha_s\Gamma_1^t}\right),
\end{eqnarray}
where $\sigma_0$ is  the LO contribution to the $tH$ associated production rate, $\Gamma_0^t, \Gamma_0^H$ are the LO  total top quark and Higgs boson decay width, and $\delta\Gamma_0^t, \delta\Gamma_0^H$ are the decay width of the top quark decaying into $\nu_{l} l^+ b$ and the Higgs boson decaying into $b\bar{b}$. $\alpha_s\sigma_1, \alpha_s\Gamma_1^i$ and $\alpha_s \delta\Gamma_1^i$, with $i=t,H$, are the corresponding NLO corrections.
We choose $\Gamma^H_{0}=3.28\times10^{-3} \textrm{~GeV}$,  $\alpha_s\Gamma^H_{1}=4.07\times10^{-3} \textrm{~GeV}$. And we calculate width of the Higgs boson by Bridge at the LO~\cite{Meade:2007js}, and adopt NLO results in Ref.~\cite{Denner:2011mq}. Here we use the modified narrow width approximation (MNW) incorporating the finite width effects as the treatment in Refs.~\cite{Cao:2004ap,Campbell:2004ch}.

We expand Eq.~(\ref{pro_dec_cro}) to order $\alpha_s$,
\begin{eqnarray}\label{pro_dec_expand}
\sigma^{NLO} =&& \sigma_0\times\frac{\delta\Gamma_0^t}{\Gamma_0^t}\frac{\delta\Gamma_0^H}{\Gamma_0^H}+\alpha_s\sigma_1\times\frac{\delta\Gamma_0^t}{\Gamma_0^t}\frac{\delta\Gamma_0^H}{\Gamma_0^H}
+\sigma_0\times\frac{\alpha_s \delta\Gamma_1^t}{\Gamma_0^t}\frac{\delta\Gamma_0^H}{\Gamma_0^H} - \nonumber\\
\nonumber\\
&&  \sigma_0\times\frac{\delta\Gamma_0^t}{\Gamma_0^t}\frac{\alpha_s\Gamma_1^t}{\Gamma_0^t}\frac{\delta\Gamma_0^H}
{\Gamma_0^H} + \sigma_0\times\frac{\alpha_s \delta\Gamma_1^H}{\Gamma_0^H}\frac{\delta\Gamma_0^t}{\Gamma_0^t} - \sigma_0\times\frac{\delta\Gamma_0^H}{\Gamma_0^H}\frac{\alpha_s\Gamma_1^H}{\Gamma_0^H}\frac{\delta\Gamma_0^t}
{\Gamma_0^t} + \mathcal {O}(\alpha_s^2).
\end{eqnarray}
Now we can separate QCD NLO  corrections  into three classes, i.e., the $tH$ associated production at the NLO with subsequent decay at the  LO, production at the LO with subsequent decay at the  NLO, and production and decay at the LO but having NLO corrections from MNW. We note that QCD NLO  corrections to the top quark decay part will contribute little since the branching ratio of the top quark  decaying into $W$ boson is always $100\%$~\cite{Gao:2011fx}. As a consequence, the third term and the fourth term of Eq.~(\ref{pro_dec_expand}) almost cancel each other. Since we only consider one decay mode of Higgs bosons, the sum of the fifth term and the sixth term will give a negative contribution.

There is another process, which is at the same  order as the $tH$ associated production at the NLO (Fig.~\ref{f_signature} (Right)), i.e., $t\bar{t}$ production with the top quark semilepton decay and the antitop decaying into $H\bar{q}$ via FCNC vertex. It has the same signal if the light quark from the top quark is missed by the detector. This additional contribution is very significant for detecting the $g_{ct}$ couplings, because the $cg \rightarrow tH$ process is suppressed by $c$ quark PDF.

\subsection{Collider simulation}
To account for the resolution of the detectors, we apply energy and momentum smearing effects to the final states~\cite{Aad:2009wy}:
\begin{eqnarray}\label{E_smear}
&&\Delta E_l/E_l = 0.05/\sqrt{E_l/{\rm GeV}}\oplus0.0055,\nonumber\\
&&\Delta E_{b(j)}/E_{b(j)} = 1/\sqrt{E_{b(j)}/{\rm GeV}}\oplus0.05,
\end{eqnarray}
where $E_{l,b,j}$ are the energy of the lepton, $b$ jet and the other jets, respectively.

We use the anti-$k_t$ jet algorithm~\cite{Cacciari:2008gp} with the jet radius $R=0.4$ and require the final-state particle to satisfy the following basic kinematic cuts
\begin{eqnarray}
&&p_{T_l}>25 {\rm ~GeV}, \quad E\!\!\!\!\slash_{T}>25 {\rm~GeV}, \quad p_{T_{b,j}}>25 {\rm~GeV},\nonumber \\
&&|\eta_l|<2.5, \quad |\eta_{b,j}|<2.5, \quad \Delta R_{bl,jl,bj}>0.4,
\end{eqnarray}
Here $E\!\!\!\!\slash_{T}$ is the missing transverse energy. $p_{T_{b,j,l}}$ and $\eta_{b,j,l}$ are the transverse momentum and pseudorapidity of the $b$ jet, other quark jets and leptons, respectively.
And $\Delta R=\sqrt{(\Delta\eta)^2+(\Delta\phi)^2}$ stands for the angular distance. Moreover, we choose a $b$-tagging efficiency of 0.6 for $b$ jets and  mistagging rates  of  1\% for other quarks.

\subsection{Events selection}

The main  background arises from the $t\bar{t}$ production with one top quark leptonic decay and the other top quark hadronic decay. Other backgrounds include $Wbbj$, $WZj$  and $Wjjj$ production processes. These backgrounds are calculated at the LO by using the MADGRAPH4~\cite{Alwall:2007st} and ALPGEN~\cite{Mangano:2002ea} programme.

For the signal, we require three tagged $b$ jets $b^{1,2,3}$, a lepton $l^+$ and the missing energy in the final states at both LO and NLO. When considering the number and kind of the final-state jets after jet clustering, there are several cases as follows:

(1) Fewer than three exclusive $b$ jets where two $b$ jets are combined together. We discard such events.

(2) Three exclusive $b$ jets. It can be from the LO and  virtual corrections, which have three bottom quarks in the final states. The final states of real corrections can also give three $b$ jets if the additional emitted parton is combined into one of the $b$ jets.  As a result, some combination procedure of jets may be different from the real partonic process. For example, the additional gluon comes from the initial states or the top quark decay, but it is combined into the $b$ jets arising from the Higgs boson, which will change the reconstructed particle mass spectrum.

(3) Four exclusive jets where the additional jet comes from a gluon or light quark. We only use the three tagged $b$ jets, neglecting the additional jet, in the final states to reconstruct the top quark and the Higgs boson. If the additional jet comes from the decay process, the reconstructed particle mass will be lower than the exact value due to lack of the momentum of the additional jet.

(4) Four exclusive jets where the additional jet comes from a $b(\bar{b})$ quark. If all of these jets are tagged, we must distinguish which one is the additional jet. But the cross section for such process is strongly suppressed by the $b(\bar{b})$ quark PDF, this contribution is neglectable.

The momentum of neutrino can be obtained by solving the on shell mass-energy equation of the $W$ boson
\begin{eqnarray}\label{w_mass_energy}
  (p_l+p_{\nu,T}+p_{\nu,L})^{2} = m_{W,r}^{2},
\end{eqnarray}
where $p_l$ is the momentum of the lepton, and $m_{W,r}$ is the reconstructed $W$ boson mass. We denote the longitudinal and transverse momentum of the neutrino as $p_{\nu,L}$ and $p_{\nu,T}$, respectively.
Since we generate intermediated particles with the MNW method, the masses of the intermediated particles  change with Breit-Weinger distribution in a region of twenty times of their widthes around the central value set in Eq.~(\ref{sm_para}). At the same time, smearing effects on the final-state particles also  affect the reconstructed $W$ mass.  As a result, if we choose a low mass of $W$ boson, there  may be no solution for $p_{\nu,L}$ in Eq.~(\ref{w_mass_energy}). So we must estimate a $W$ boson mass, which is large enough to provide the solution of mass-energy equation and small enough to reconstruct the proper top quark mass.  Since there are three $b$ jets in the final states, it is necessary to determine the proper combination of the $b$ jets to reconstruct the top quark mass $m_{t,r}$ and the Higgs boson mass $m_{H,r}$.  In practice, we adopt the following steps:

(1) We choose  $m_{W,r}$ randomly with Breit-Weinger distribution when solving Eq.~(\ref{w_mass_energy}) to get the neutrino longitudinal momentum. For such $m_{W,r}$, it is required to  provide real solutions of the equation.  If not, discard this value of $m_{W,r}$ and redo this step until we obtain  real solutions.

(2) In order to improve the $W$ boson reconstruction efficiency and  save calculation time, we repeat step 1 for ten times and denote the solutions as $p_{\nu,L}^{i,j}$, with $ i=1,2, j=1, \cdot\cdot\cdot, 10$. The superscript $i$ represents the number of the solutions  in Eq.~(\ref{w_mass_energy}).

(3) Choose one momentum of the $b$ jets  $p_b^{k}, k=1,2,3$  and one of  $p_{\nu,L}^{i,j}$ to reconstruct the top quark mass $m_{t,r}^{i,j,k}=\sqrt{(p_l)^2+(p_{\nu,L}^{i,j})^2+(p_{\nu,T})^2+(p_b^{k})^2}$.

(4) For all combinations of $(i,j,k)$, we choose the best one to minimize  $|m_{t,r}^{i,j,k} - m_t|$.

(5) Use the remnant $b$ jets to reconstruct Higgs boson mass $m_{H,r}^{k}$.

In order to choose appropriate kinematic cuts, we show some important kinematic distributions for the signal and the backgrounds.
\begin{figure}[h]
  \includegraphics[width=0.5\linewidth]{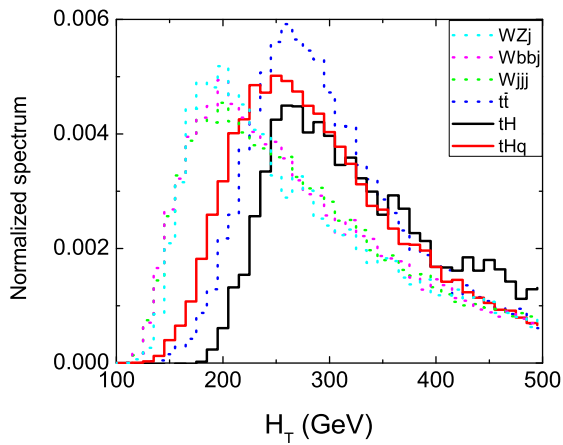}\\
  \caption{Dependence of differential cross sections on $H_T$. The label $tH$ represents the $tH$ associated production, while the label $tHq$ stands for the process of $t\bar{t}$ production with  rare decay mode $t\bar{t}\rightarrow tH\bar{q}$. The other labels denote the backgrounds.}\label{f_HT}
\end{figure}
In Fig.~\ref{f_HT}, we present dependence of differential cross sections of the signal and backgrounds on $H_T$,  defined as the scalar sum of lepton and jet transverse momenta. From the figure, we can see that the distributions of $Wjjj$, $WZj$ and $Wbbj$ backgrounds have peaks below $200$ GeV, while the peak positions of the signals are about $240$ GeV. Therefore we choose the $H_T$ cut
\begin{eqnarray}
H_T > 200  ~{\rm GeV} .
\end{eqnarray}

\begin{figure}[h]
\begin{minipage}[t]{0.45\linewidth}
\centering
  \includegraphics[width=1.1\linewidth]{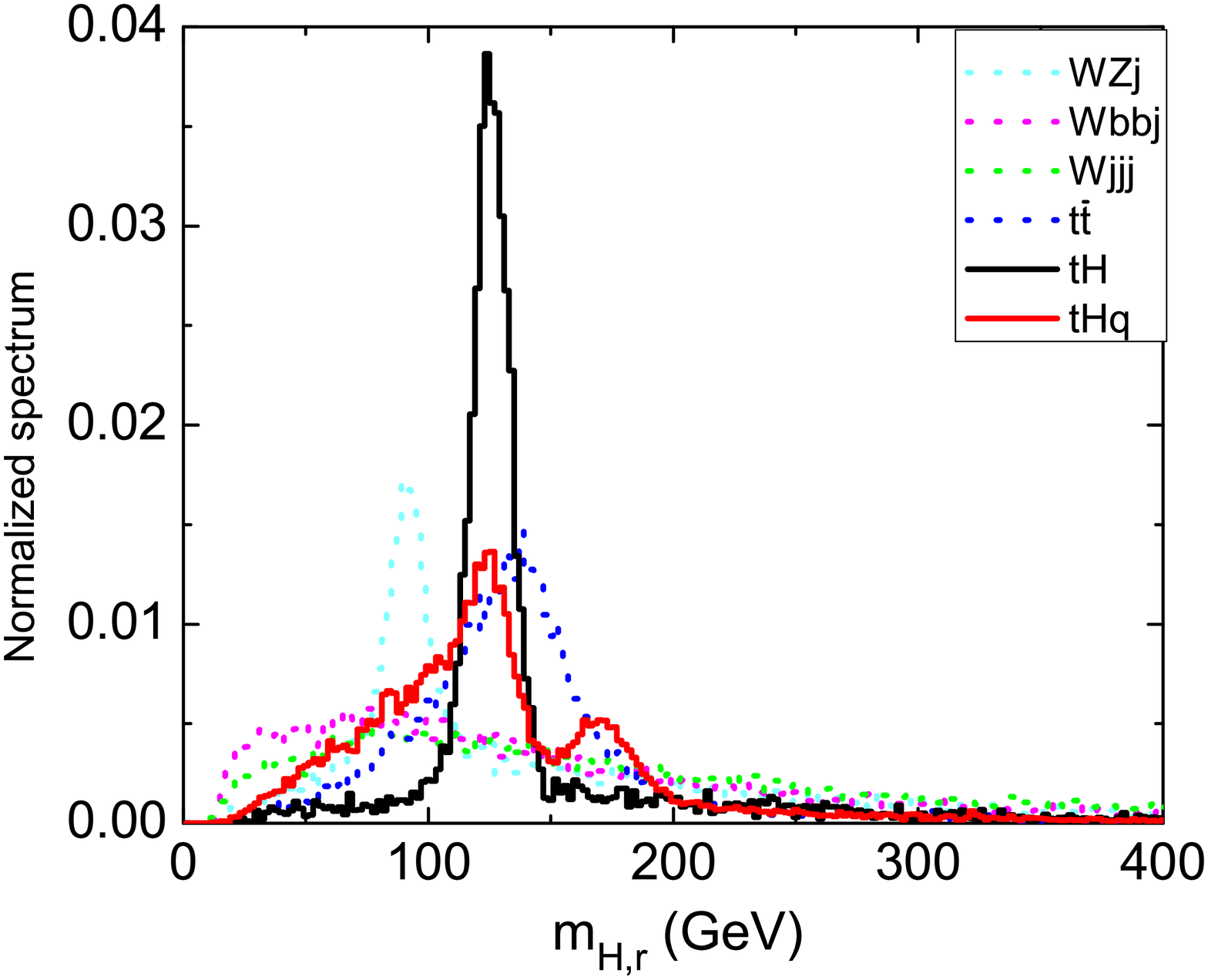}\\
\end{minipage}
\hfill
\begin{minipage}[t]{0.45\linewidth}
\centering
  \includegraphics[width=1.1\linewidth]{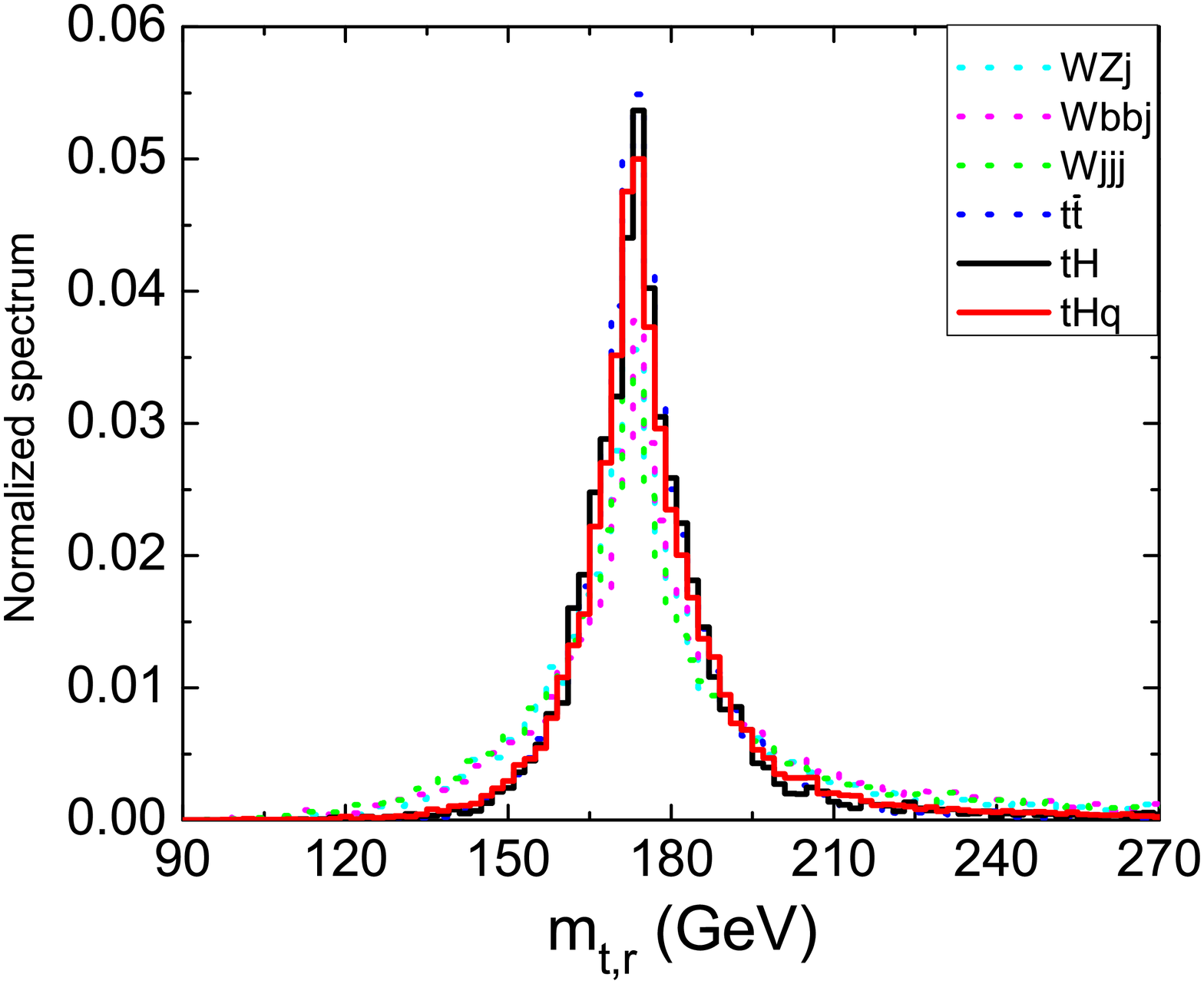}\\
\end{minipage}
\caption{The reconstructed top quark and  Higgs boson mass. The labels here are the same as those in Fig~\ref{f_HT}.}\label{f_bk_re_m}
\end{figure}
Figure~\ref{f_bk_re_m} illustrates the distribution of the reconstructed Higgs boson mass of the signal and backgrounds. We can see that the signals have peaks around 125 GeV, while the  distributions of backgrounds are continuous or have peaks at other places.  In order to suppress $Wbbj$, $WZj$ and $Wjjj$ backgrounds, we  require the mass of the Higgs boson to satisfy
\begin{eqnarray}\label{mh_cut}
\Delta m_{H,r} < 20 {\rm GeV},
\end{eqnarray}
where $\Delta m_{H,r}$ is defined as $|m_{H,r}-m_H|$.
The reason will be explained in more detail in the Appendix. We also show the distribution of the reconstructed top quark mass in Fig.~\ref{f_bk_re_m},  where the signal and the backgrounds have  similar distributions. As a result, we choose the cut
\begin{eqnarray}\label{mt_cut}
\Delta m_{t,r} < 20 {\rm GeV}
\end{eqnarray}
to keep more signal events, where $\Delta m_{t,r}$ is defined as $|m_{t,r}-m_t|$.

\begin{figure}[h]
\centering
  \includegraphics[width=0.5\linewidth]{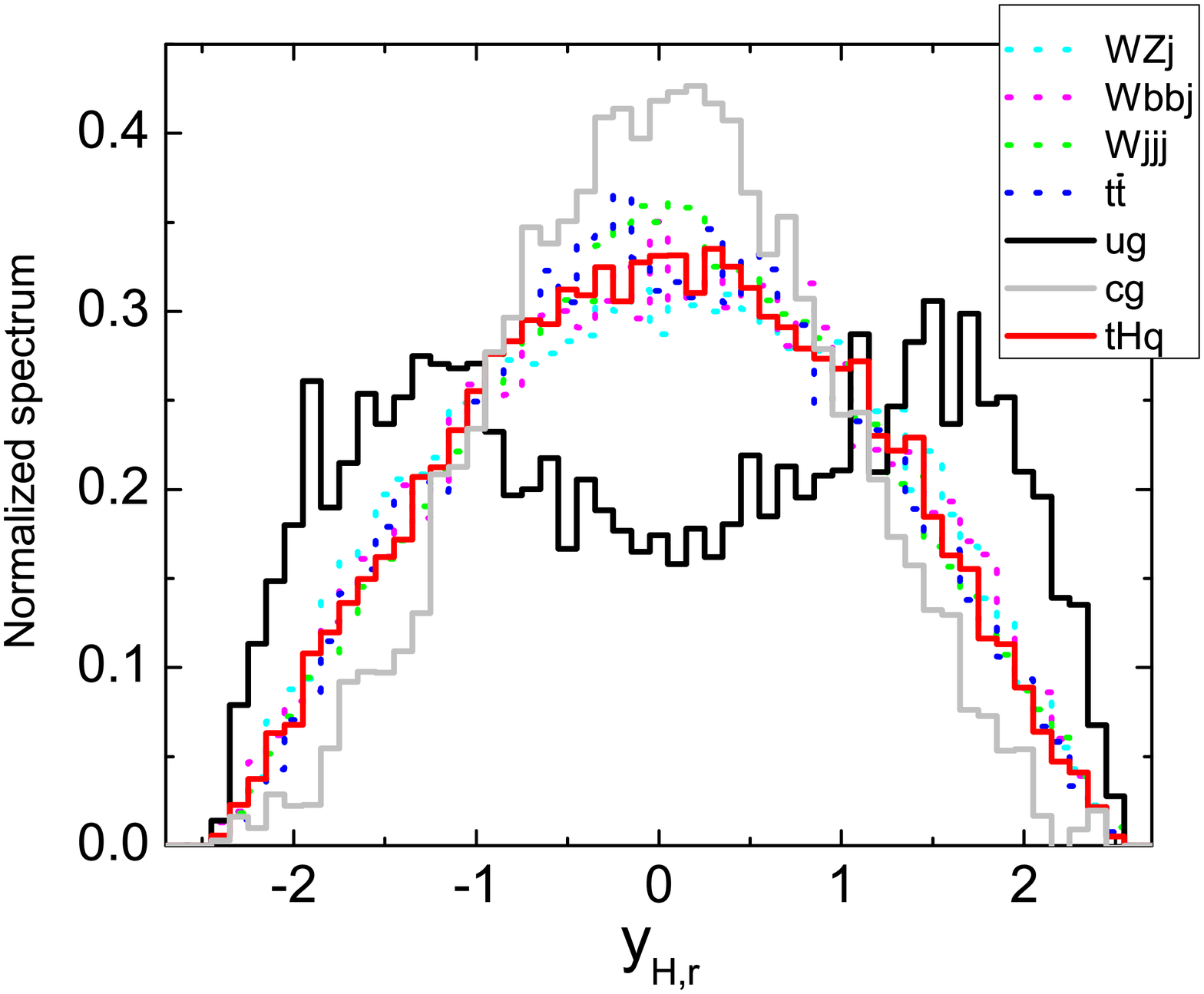}\\
\caption{Dependence of differential cross sections on the rapidity of the Higgs boson and the top quark. The labels $u(c)g$ represent the $u(c)g \rightarrow tH$ associated productions, while the label $tHq$ stands for the process of $t\bar{t}$ production with  rare decay mode $t\bar{t}\rightarrow tH\bar{q}$. The other labels denote the backgrounds.}  \label{f_bk_re_eta}
\end{figure}

To determine the rapidity cut,  we present  the normalized spectrum  of  the rapidity of the reconstructed resonances for the signal and backgrounds in Fig.~\ref{f_bk_re_eta}. It can be seen the Higgs boson from the $ug \rightarrow tH$ associated production concentrates in the forwards and backwards regions. This is due to the fact that the momentum of initial $u$ quark is generally larger than that of gluon, so the partonic center-of-mass frame is highly boosted along the direction of the $u$ quark.
On the contrary, the main contribution of top pair production comes from gluon initial-states, which are symmetric and have small boost effect.  So we impose rapidity cut on reconstructed Higgs boson for the signal of $ug \rightarrow tH$ process as
\begin{eqnarray}\label{y_cut}
|y_{H,r}|>1.0.
\end{eqnarray}
However, since $c$ quark is the sea quark, the momentum of initial $c$ quark is much smaller than that of the initial $u$ quark. As a result, the Higgs boson from $cg$ initial states is not boosted as from $ug$ initial states. In addition, the cross section of $t\bar{t}\rightarrow tH\bar{c}$ decay  is comparable to that of $cg \rightarrow tH$ process. Therefore when discussing the $Htc$ couplings, we do not apply the cut in Eq.~(\ref{y_cut}).

The  complete set of kinematical cuts is listed in Table~\ref{t_cuts}.
\begin{table}[h]
\begin{center}
\begin{tabular}{cccccccc}
  \hline
  \hline
   & basic cut  \quad \quad   && $p_{T_j} >25 $ GeV, \quad  $p_{T_l} >25 $ GeV, \quad  $\quad E\!\!\!\!\slash_{T} > 25 $ GeV, \\
   &             \quad \quad   && \quad $|\eta_{l,j}| <2.5, $\quad  $\Delta R_{bl,jl,bj}> 0.4$ \\
  \hline
   & $H_T$       \quad \quad  && $H_T > 200 $ GeV         \\
  \hline
   & $m_{t,r}$       \quad \quad  && $\Delta m_t<20$ GeV   \\
  \hline
   & $m_{H,r}$       \quad \quad  && $\Delta m_H<20$ GeV   \\ 
  \hline
   & $y_{H,r}$    \quad \quad  && $|y_{H,r}|>1.0$    \\
  \hline
  \hline
\end{tabular}
\end{center}
 \caption{Kinematic cuts in the event selection.}\label{t_cuts}
\end{table}

\subsection{Simulation results}
In this subsection, we discuss the numerical results after imposing kinematic cuts. We need to include the QCD NLO  corrections to the decay as well. As a result, we find that these corrections  reduce the cross sections by about 50\%  for  the $g_{ut}$ coupling induced process and by about $100\%$ for the $g_{ct}$ coupling induced process, respectively. The corresponding results are listed in Table~\ref{t_tot_kfactor}. $K_{pro}$ only includes the NLO corrections to the $tH$ associated production, while $K_{tot}$ also contains the NLO corrections to decay. The complete QCD NLO  corrections are $11\%$  for $ug \rightarrow tH$ and almost vanish for $cg \rightarrow tH$.
\begin{table}[h]
\begin{center}
\begin{tabular}{cccccccc}
  \hline
  \hline
    \quad \quad \quad  \quad \quad \quad& $\sigma_{LO}$ [fb]  \quad \quad \quad \quad \quad \quad & $K_{pro}$   \quad \quad \quad\quad \quad \quad & $K_{tot}$   \\
  \hline
    $tug$       &   6.64    \quad \quad \quad  \quad \quad \quad       &1.22       \quad \quad \quad   \quad \quad \quad  &  1.11        \\
  \hline
    $tcg$       &   0.428   \quad \quad \quad   \quad \quad \quad      &1.40     \quad \quad \quad   \quad \quad \quad    &  1.00         \\
  \hline
  \hline
\end{tabular}
\end{center}
 \caption{The LO cross sections and K factors of the $tH$ associated  production at the LHC ($g_{qt}=0.2$) with the kinematic cuts in Table~\ref{t_cuts} applied. We define $K_{pro}$ and $K_{tot}$ to be the K factor of the  process of $tH$ associated  production and the processes including production and decay, respectively. We apply all  cuts for the $g_{ut}$ coupling induced process. But for the $g_{ct}$ coupling induced process, we do not apply the rapidity cut on the reconstructed Higgs boson.}\label{t_tot_kfactor}
\end{table}

\begin{table}[h]
\begin{center}
\begin{tabular}{cccccccc}
  \hline
  \hline
    \quad \quad \quad             & basic cut           \quad \quad \quad & $H_T$   \quad \quad \quad &      $m_{t,r}$         \quad \quad \quad & $m_{H,r}$       \quad \quad \quad &  $y_{H,r}$     \quad \quad \quad & $\epsilon_{cut}$ \\
  \hline
   $ug\rightarrow tH$(NLO)        &   3.54              \quad \quad \quad &3.26     \quad \quad \quad &      2.86          \quad \quad \quad &2.22         \quad \quad \quad &   1.59    \quad \quad \quad & 44.9\%     \\
  \hline
   $cg\rightarrow tH$(NLO)        &   0.40             \quad \quad \quad &0.354    \quad \quad \quad &      0.322         \quad \quad \quad &0.240        \quad \quad \quad &   0.09    \quad \quad \quad &  22.5\%     \\
  \hline
    $t\bar{t}\rightarrow tH\bar{q}$(LO)                     &   0.993             \quad \quad \quad &0.956    \quad \quad \quad &      0.849         \quad \quad \quad &0.487       \quad \quad \quad &   0.193    \quad \quad \quad & 19.4\%     \\
  \hline
   $t\bar{t}$ \quad               &   21.0              \quad \quad \quad &19.9     \quad \quad \quad &      17.3          \quad \quad \quad &9.55         \quad \quad \quad &   3.89     \quad \quad \quad &  18.5\%     \\
  \hline
   $W^{+}b\bar{b}j$ \quad         &   3.30              \quad \quad \quad &2.32     \quad \quad \quad &      1.38         \quad \quad \quad &0.336       \quad \quad \quad &   0.146    \quad \quad \quad &   4.4\%     \\
  \hline
  $W^{+}Zj$\quad                  &   0.215             \quad \quad \quad &0.160    \quad \quad \quad &      0.099         \quad \quad \quad &0.023        \quad \quad \quad &   0.010    \quad \quad \quad &    4.7\%     \\
  \hline
   $W^{+}jjj$                      &   0.085             \quad \quad \quad &0.064    \quad \quad \quad &      0.038         \quad \quad \quad &0.009        \quad \quad \quad &   0.004    \quad \quad \quad &   4.7\%     \\
  \hline
  \hline
\end{tabular}
\end{center}
 \caption{Cross sections (in {\rm fb}) after imposing cuts for NLO $tH$, $tH(\bar{q})$ signal, and their backgrounds $t\bar{t}$, $Wb\bar{b}j$, $Wjjj$ and $WZj$. The cut acceptance $\epsilon_{cut}$ is also listed.  The $b$-tagging efficiency has been taken into account in the basic cut.} \label{t_events_loose}
\end{table}

We list the results after imposing various kinematic cuts in Table ~\ref{t_events_loose}. For the $ug \rightarrow tH$ process, the clear signal with the coupling $g_{ut} = 0.2$ can be observed at the $5\sigma$ C.L. when the integrated luminosity is $100~{\rm fb}^{-1}$ at the LHC. Here we define the discovery significance as $\mathcal {S} /\sqrt{\mathcal {B}} =5$ and exclusion limits as $\mathcal {S}/\sqrt{\mathcal {S}+\mathcal {B}} = 3$, where $\mathcal {S}$ and $\mathcal {B}$ are the expected events numbers of the signal and the backgrounds. However, for the $ug \rightarrow tH$ process, the cross section of the process $t\bar{t} \rightarrow tH\bar{c}$ is about 2 times larger than that of the process $cg \rightarrow tH$ after cuts. As stated before, we choose data from the fifth column of Table~\ref{t_events_loose}. As a result, the $5\sigma$ C.L. discovery sensitivity of $g_{ct}$ is 0.294 when the integrated luminosity is $100~{\rm fb}^{-1}$ and $m_H =125$ GeV.

\begin{figure}[h]
\begin{minipage}[t]{0.45\linewidth}
\centering
  \includegraphics[width=1.1\linewidth]{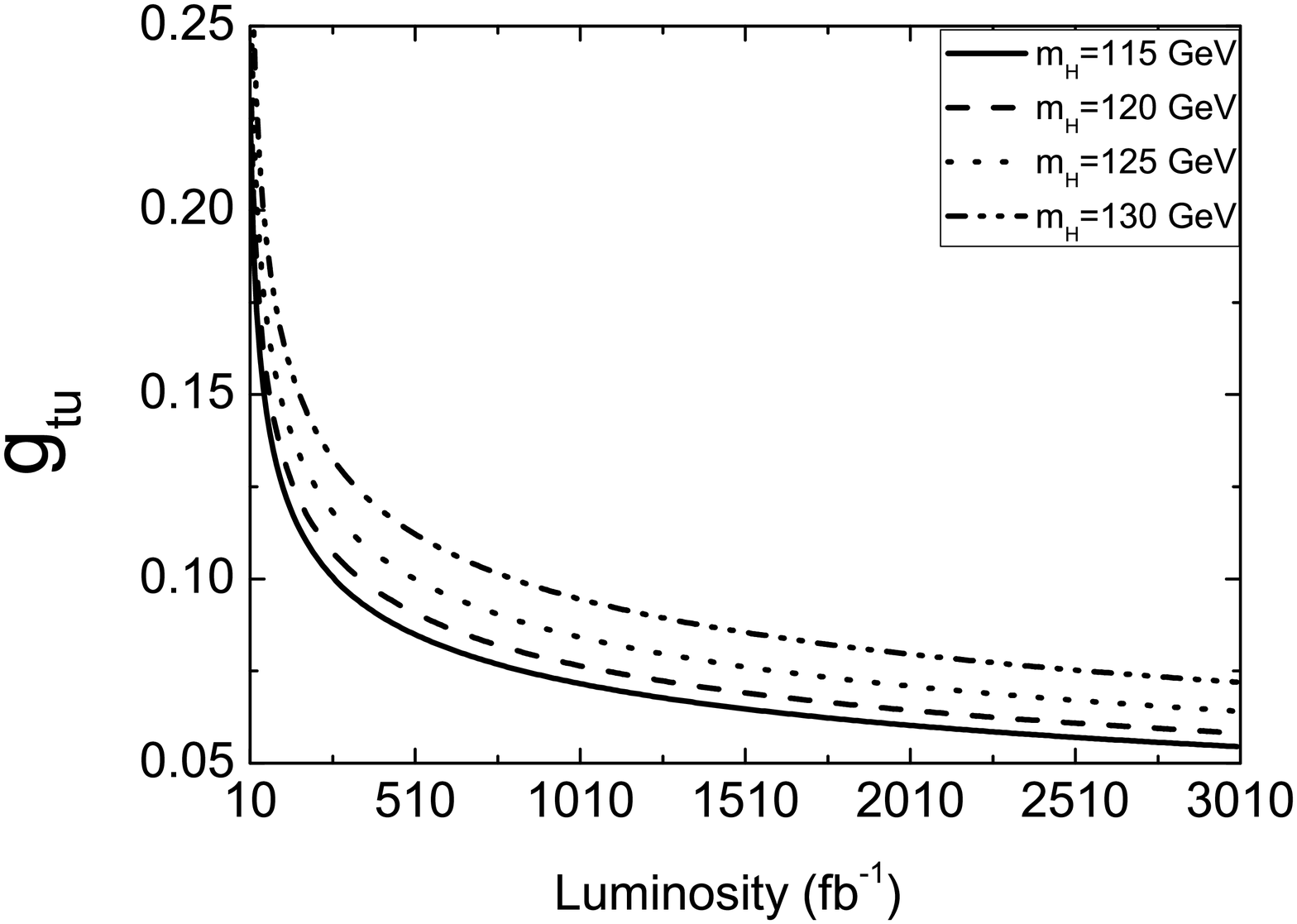}\\
\end{minipage}
\hfill
\begin{minipage}[t]{0.45\linewidth}
\centering
  \includegraphics[width=1.1\linewidth]{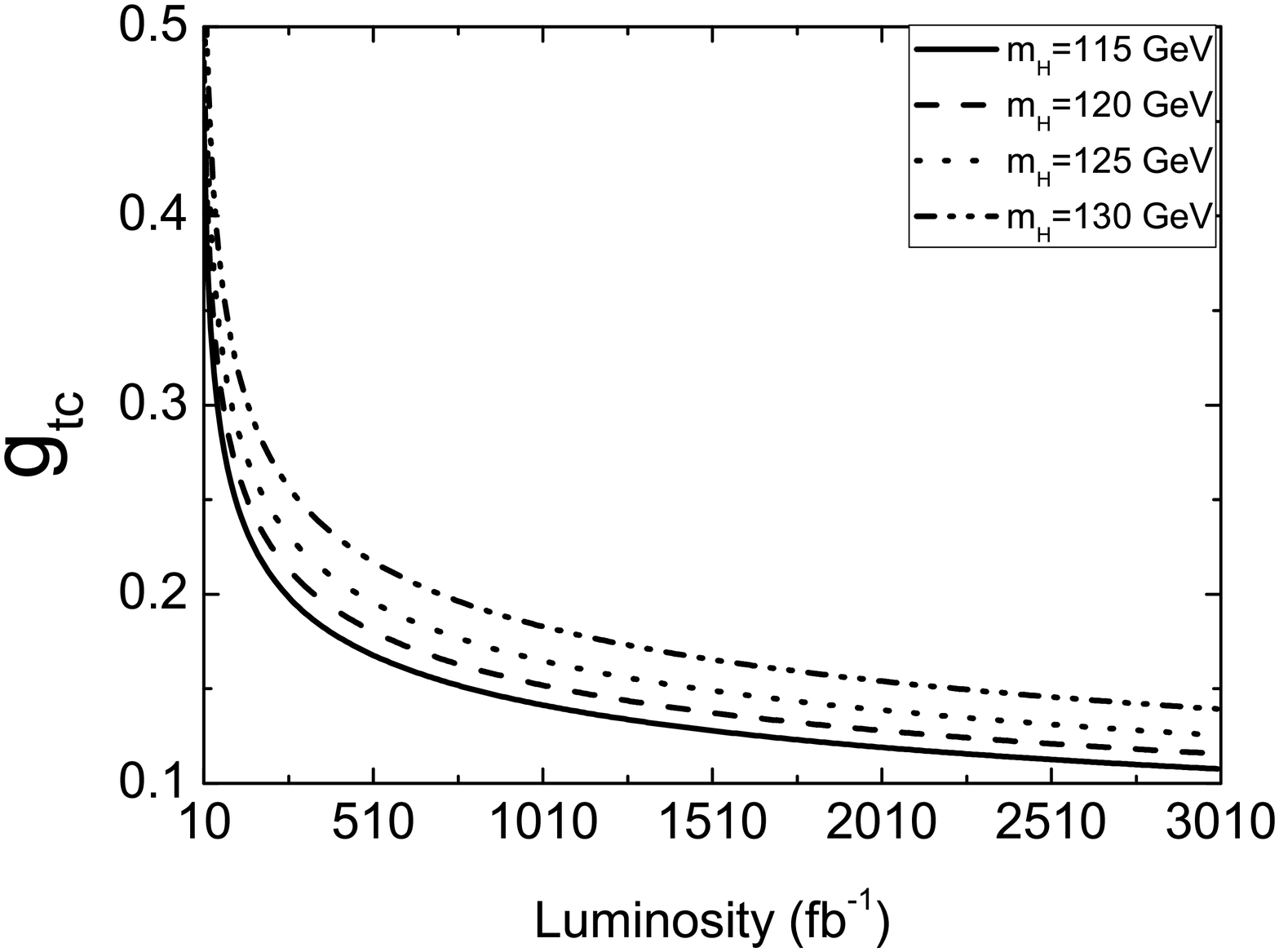}\\
\end{minipage}
\caption{The 5$\sigma$ C.L. discovery sensitivities to the FCNC $Htq$ couplings. The lines from bottom to top correspond to the $m_H$ from $115$ GeV to $130$ GeV.}\label{f_coup_5sigma}
\end{figure}
We show the $5\sigma$ discovery sensitivities  to FCNC couplings with several Higgs boson mass for different luminosities in Fig.~\ref{f_coup_5sigma}.  For a lighter Higgs boson, the cross sections become larger, but the branching ratio rates of $Br(H \rightarrow b\bar{b})$ get lower. When $m_H =115$ GeV and integrated luminosity is $100~{\rm fb}^{-1}$, the limit on the $g_{ut}$ coupling is 15\% smaller than that for  $m_H =125$ GeV. In contrast, when the Higgs boson mass increases to $m_H=130$ GeV, the limit is increased by 12.3\%. The $g_{ct}$ coupling has the similar behavior.

\begin{figure}[h]
\begin{minipage}[t]{0.45\linewidth}
\centering
  \includegraphics[width=1.1\linewidth]{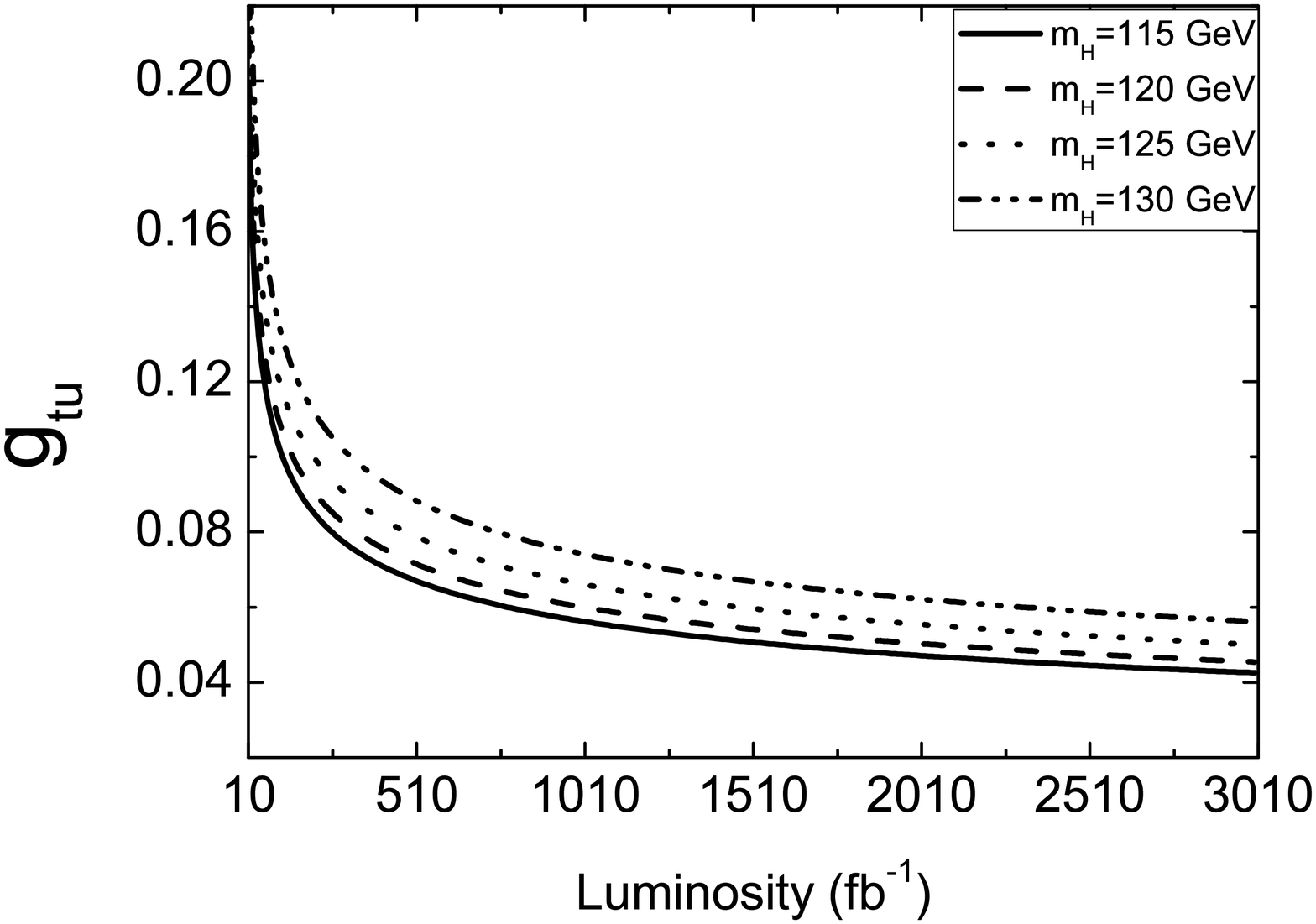}\\
\end{minipage}
\hfill
\begin{minipage}[t]{0.45\linewidth}
\centering
  \includegraphics[width=1.1\linewidth]{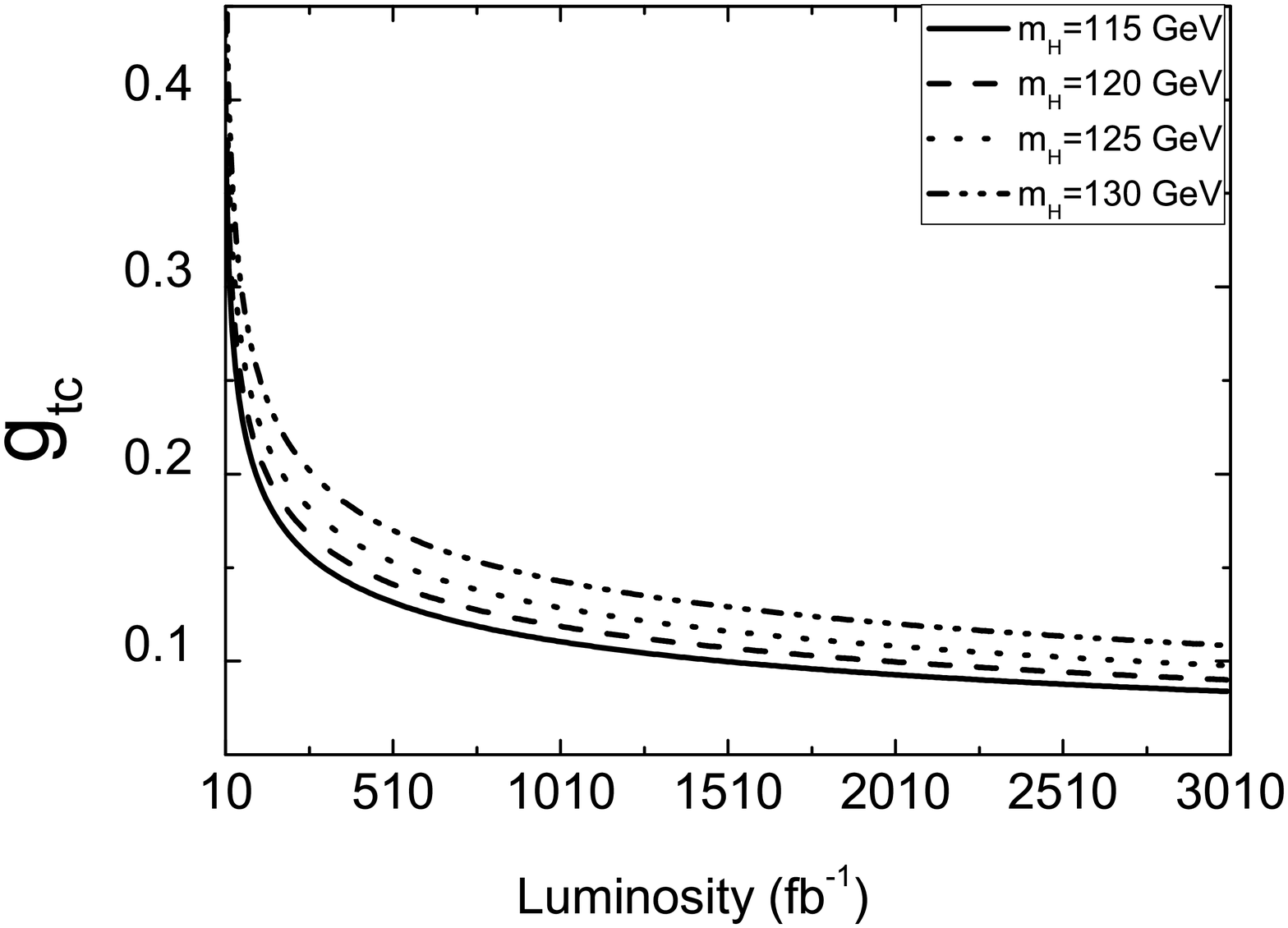}\\
\end{minipage}
\caption{The 3$\sigma$ C.L. exclusion limits to the $Htq$ couplings. The lines from bottom to top correspond to the $m_H$ from $115$ GeV to $130$ GeV.}\label{f_coup_3sigma}
\end{figure}
If no signal is observed, it means that the FCNC $Htq$ couplings can not be too large. In Fig.~\ref{f_coup_3sigma},  we show the $3\sigma$ exclusion limits of couplings with several Higgs boson masses for different luminosities. The upper limits on the size of FCNC couplings are given as $g_{ut}\leq 0.121$ and  $g_{ct}\leq 0.233$ with $m_H =125$ GeV. These limits can be converted to the $3\sigma$ C.L. upper limits on the branching ratios of top quark rare decays~\cite{Li:1990qf,AguilarSaavedra:2004wm} as follows:
\begin{eqnarray}
Br(t\rightarrow Hu) \leq 4.1\times10^{-4}, \quad Br(t\rightarrow Hc) \leq 1.5\times10^{-3}.
\end{eqnarray}

\section{CONCLUSION}\label{s6}
In conclusion, we have investigated the signal of the $tH$ associated production via the FCNC $Htq$ couplings at the LHC with $\sqrt{S}= 14$ TeV, including complete QCD NLO  corrections to the production and decay of top quark and Higgs boson. Our results show that the NLO corrections reduce the scale dependences of the total cross sections, and increase the production cross sections by $48.9\%$ and $57.9\%$ for the $Htu$  and $Htc$  couplings induced processes, respectively. After kinematic cuts are imposed on the decay products of the top quark and the Higgs boson, the NLO corrections are reduced to $11\%$  for the $Htu$ coupling induced process and almost vanish for the $Htc$ coupling induced process. For the signal, we discuss the Monte Carlo simulation results for the signal and corresponding backgrounds, including the process of top quark pair production with one of the top quarks decaying to $Hq$ as well, and  show that the NP signals may be observed at the $5\sigma$ level in some parameter regions. Otherwise, the $3\sigma$ upper limits on the FCNC couplings can be set, which can be converted to  the constraints on the top quark rare decay branching ratios.

\section{Acknowledgements}
This work was supported by the National Natural
Science Foundation of China, under Grants
No. 11021092, No. 10975004 and No. 11135003.

\section*{APPENDIX}
In this appendix,  we  numerically check  the mass cut of the reconstructed particles.
As stated before, the emission of an extra gluon  broadens the mass distributions of reconstructed particles and makes $b$-jet softer, which  decreases the K factor of QCD NLO  corrections when imposing  reconstructing mass cuts or $b$-jet $p_t$ cuts. The more strict mass cuts are imposed, the smaller cross sections we get. On the contrary,  if the mass cuts are loose, though the cross sections  are larger, more background events are also be considered, which may decrease the signal to background ratio. As a result, it is difficult to choose mass cuts on the reconstructed particles. We have checked that when we change $\Delta m_H $ from $< 5$ GeV to $< 25$ GeV,  the sensitivity to the $g_{ut}$ coupling at the $5\sigma$ level is the lowest when $\Delta m_H <20$ GeV, as shown in Table~\ref{t_mh_coupling}.  It confirms  our choice in Eq.~(\ref{mh_cut}).
\begin{table}[h]
\begin{center}
\begin{tabular}{cccccccc}
  \hline
  \hline
   $\Delta m_H$               & $<5$ GeV            & $<10$ GeV   &   $ <  15$ GeV        & $<20$ GeV        &  $<25$ GeV \\
  \hline
   sensitivity to $g_{ut}$        &   0.180           &0.159    &      0.157        & 0.150  &   0.335\\
  \hline
  \hline
\end{tabular}
\end{center}
 \caption{Behavior of the sensitivity to the FCNC $g_{ut}$  couplings as a function of $\Delta m_H$ cuts at the $5\sigma$ level.}\label{t_mh_coupling}
\end{table}

\newpage

\bibliography{tqv}
\end{document}